\documentclass[lettersize, journal]{IEEEtran}

\usepackage{amsfonts}
\usepackage{amsmath}
\usepackage{amsmath,epsfig,amsmath}
\usepackage{amsfonts,amssymb,amsthm}
\usepackage{epsfig}
\usepackage{float}
\usepackage{graphicx}
\usepackage{color,hyperref}
\usepackage{multirow}
\usepackage{tipa}
\usepackage{picins}
\usepackage{color}
\usepackage{microtype}
\usepackage{booktabs}
\usepackage{stfloats}
\usepackage{bigstrut,multirow,rotating,booktabs}
\usepackage[utf8]{inputenc}
\usepackage[T1]{fontenc}
\usepackage[table]{xcolor}
\usepackage{threeparttable}
\usepackage[lined,boxed,commentsnumbered,ruled]{algorithm2e}

\begin{document}

\title{EQ-TAA: Equivariant Traffic Accident Anticipation via Diffusion-Based Accident Video Synthesis}

\author{Jianwu Fang, Lei-Lei Li, Zhedong Zheng, Hongkai Yu,  Jianru Xue,  Zhengguo Li, and Tat-Seng Chua
\thanks{J. Fang, L. Li, and J. Xue are with the Xi'an Jiaotong University, Xi'an, China
     {(fangjianwu@mail.xjtu.edu.cn).}}
\thanks{Z. Zheng is with the University of Macau, China {(zhedongzheng@um.edu.mo)}.}     
\thanks{T. Chua is with the National University of Singapore, Singapore.}
 \thanks{H. Yu is with the Department of Electrical Engineering and Computer Science, Cleveland State University, Cleveland, USA.}
\thanks{Z. Li is with the Institute for Infocomm Research, Agency for Science, Technology and Research (A$^*$STAR), Singapore.}

}

\markboth{Journal of Latex}%
{Shell \MakeLowercase{\textit{et al.}}: Bare Demo of IEEEtran.cls for Computer Society Journals}

%



\IEEEtitleabstractindextext{%
\begin{abstract}
Traffic Accident Anticipation (TAA) in traffic scenes is a challenging problem for achieving zero fatalities in the future. Current approaches typically treat TAA as a supervised learning task needing the laborious annotation of accident occurrence duration. However, the inherent long-tailed, uncertain, and fast-evolving nature of traffic scenes has the problem that real causal parts of accidents are difficult to identify and are easily dominated by data bias, resulting in a background confounding issue. Thus, we propose an Attentive Video Diffusion (AVD) model that synthesizes additional accident video clips by generating the causal part in dashcam videos, \emph{i.e.}, from normal clips to accident clips. AVD aims to generate causal video frames based on accident or accident-free text prompts while preserving the style and content of frames for TAA after video generation. This approach can be trained using datasets collected from various driving scenes without any extra annotations. Additionally, AVD facilitates an Equivariant TAA (EQ-TAA) with an equivariant triple loss for an anchor accident-free video clip, along with the generated pair of contrastive \emph{pseudo-normal} and \emph{pseudo-accident} clips. Extensive experiments have been conducted to evaluate the performance of AVD and EQ-TAA, and competitive performance compared to state-of-the-art methods has been obtained.
\end{abstract}

\begin{IEEEkeywords}
Traffic accident anticipation, video diffusion model, causal reasoning, equivariant learning, dashcam videos
\vspace{3em}
\end{IEEEkeywords}}

\maketitle

\IEEEdisplaynontitleabstractindextext

%
\IEEEpeerreviewmaketitle

\IEEEraisesectionheading{
\section{Introduction}
\label{section1}}
\IEEEPARstart{V}{ast} amount of road accidents have deprived many human lives each year. More than half of the deaths fall on vulnerable road users (\emph{i.e.,} pedestrians, cyclists, and motorbikes), and road accidents are the leading killer for young people aged from 5 to 29 years old \cite{trafficdeath}. The cause of a traffic accident is significantly influenced by the car to the road network. This shocking fact motivates many researchers and institutes to devote efforts to traffic accident detection or anticipation systems \cite{DBLP:journals/tits/ChiaKGJ22}. 

\begin{figure}[!t]
  \centering
 \includegraphics[width=\hsize]{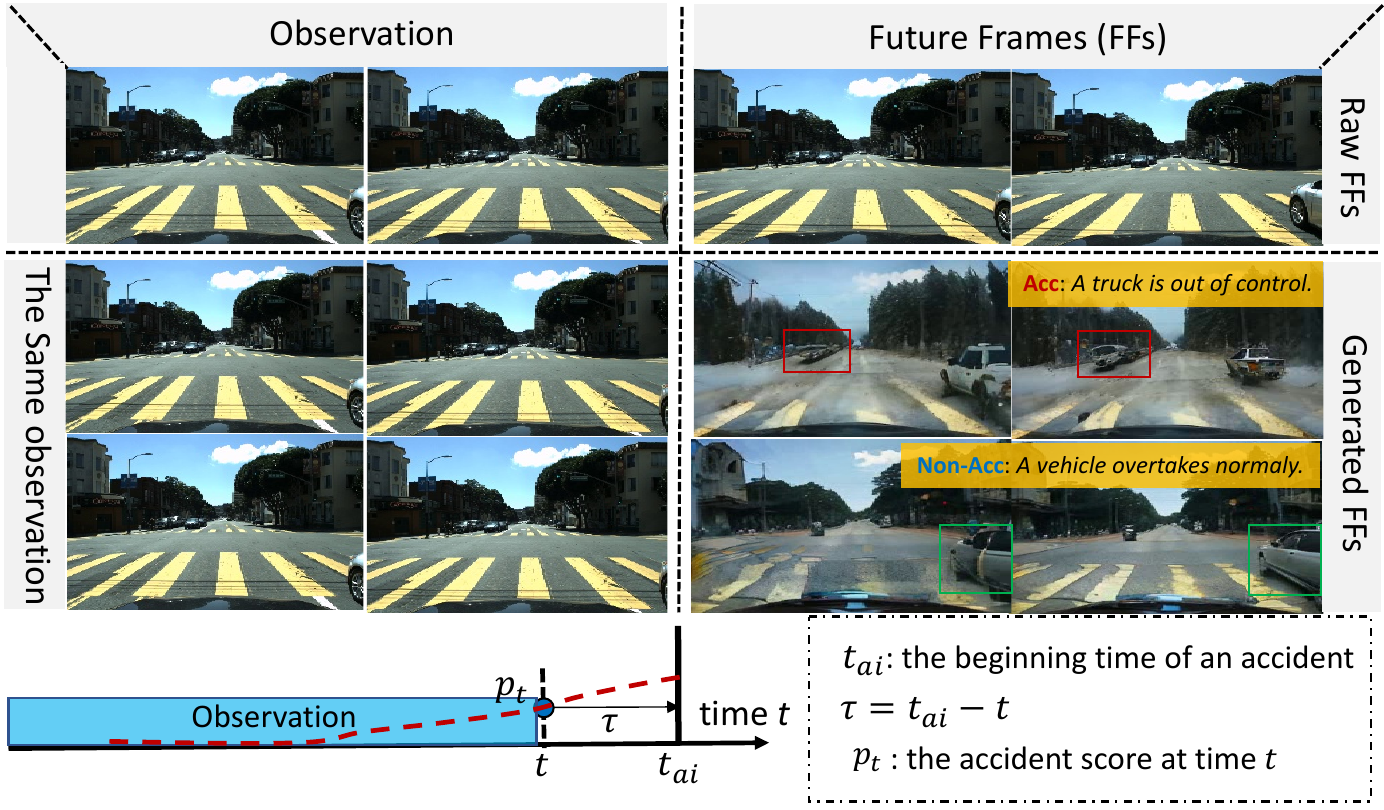}
  \caption{\small{The generated video frames by the Attentive Video Diffusion model for an accident (red boxes mark accident objects) and an accident-free (marked by green boxes) scenario. The raw videos are sampled in an accident-free HEV-I dataset \cite{yao2019egocentric}.}}
  \label{fig1}
  \vspace{-1em}
\end{figure}

In this work, we focus on the Traffic Accident Anticipation (TAA) problem which is different from the Traffic Accident Detection (TAD) \cite{fang2022traffic,DBLP:journals/corr/abs-2308-15985} with the localization of the accident window. TAA has several historical frames up to time $T$ as the observation and prefers an early accident warning for a time $t$ by maximizing a Time-to-Accident (TTA) duration $\tau$ to the starting time of accident $t_{ai}$, as illustrated in Fig. \ref{fig1}.

No one road user wants to be involved in a traffic accident. This expectation of the whole human world forms the rarity of traffic accidents in terms of time duration and spatial locations. Usually, an accident occurs suddenly and may cover only about 10-30 video frames (0.5-1 seconds), where the road users (pedestrians, cyclists, cars, etc.) take up a small spatial ratio of the whole frame. Therefore, the normal background takes up a large part of video data, which implies a confounding issue \cite{DBLP:journals/tcsv/ChenLZ23} for the causal part of accidents. 
Further, current paradigms typically formulate TAA as a supervised learning framework with the given accident time window and occurrence annotation. This limits their practical generalization as there is limited annotated data. Besides, a huge part of TAA works \cite{karim2023attention,DBLP:journals/tits/ChiaKGJ22} concentrate on the object-centric (\emph{detection$\rightarrow$ tracking}) temporal consistency or spatial interaction framework for feature extraction. However, the complex traffic scenes make these frameworks inadequate for satisfactory TAA performance. 

Some attempts take the counterfactual sample synthesis to overcome the confounding issue in video understanding tasks, \emph{e.g.},  masking~\cite{DBLP:journals/tmm/QinZHLS23,DBLP:journals/pami/ChenZNZX23}, or object/scene-level grounding~\cite{DBLP:journals/pami/WangGX23,li2022invariant} to cut off the noisy context for task prediction. However, these formulations still rely on the pre-annotated positive/negative labels. 
Recently, video diffusion models aim to generate the expected video clips with the guidance of words with clear semantic descriptions \cite{DBLP:journals/corr/abs-2303-12688,DBLP:journals/corr/abs-2303-04761,DBLP:journals/tmm/LiWFZW20}. This formulation implies a clear semantic alignment to activate the correlation between the key visual concept and given text prompts. By this feature, diffusion models can flexibly create positive and negative samples. 
Hence, for the TAA problem, we propose an Attentive Video Diffusion (AVD) model for ego-view accident video generation. The first aim of AVD is to tackle the confounding issue by generating and changing the causal part of the accident videos, \emph{i.e.}, interventing the anchor \emph{normal clip} into \emph{accident clip} explicitly. The second aim of AVD is to relax the demand for the annotation of accident video data. 

To fulfill the above two aims, the AVD needs two kinds of abilities: \textbf{1)} the changing of the causal part of a video conditioned by text prompts with a well-aligned text-video semantic learning, and \textbf{2)} preservation of the content details and temporal dependence in the generated frames owning to the problem setting of TAA. To guarantee the first ability, we adopt the stable diffusion model~\cite{DBLP:journals/corr/abs-2212-11565} as the backbone and fine-tune our AVD on a large-scale multi-modality accident dataset  \cite{DBLP:journals/corr/abs-2212-09381} with over 10k accident videos and well-aligned text-video annotation. A cross-attention model is designed in AVD to maintain the text-video semantic alignment. For the second ability, we add several frames in the original videos as the condition to the reverse diffusion process of AVD and formulate a spatial-temporal attention model to preserve the content consistency and temporal dependence in the generated video frames. This strategy also facilitates an \emph{equivariant triple loss} to contrast the causal part learning by the triple clip set (\textbf{\emph{e}-triple$_c$-set}), \emph{i.e.}, \emph{\{anchor video clip, pseudo-normal clip,  pseudo-accident clip}\}. This triple set construction has a similar consideration in the adversarial perturbation of the object Re-ID task~\cite{DBLP:journals/tomccap/LiuWHGSBB23} or contrastive learning for the online video instance segmentation~\cite{DBLP:conf/iccv/YingZMWCWLFZS23}. However, AVD promotes an Equivariant TAA (EQ-TAA), and the datasets for training EQ-TAA can be collected from arbitrary driving scenes without annotation, avoiding the laborious annotation for positive/negative samples of traditional TAAs. The \textbf{contributions} are threefold.

\begin{itemize}
\item We propose an attentive video diffusion model that can handle the confounding issue of background frames in TAA and facilitate an \emph{equivariant triple loss} in TAA to enhance the causal part learning on the \textbf{\emph{e}-triple$_c$-sets} \footnote{Each triple clip set contains an anchor video clip, a generated accident clip, and an accident-free clip.}.
\item To the best knowledge, we facilitate an Equivariant TAA (EQ-TAA) model that can be trained by the ego-view traffic videos collected from various driving scenes without any annotation for flexible generalization.
\item We conduct extensive experiments on video diffusion and TAA, and show competitive performance to other state-of-the-art TAA methods. We evaluate the video diffusion performance on the HEV-I \cite{yao2019egocentric} and BDDA dataset \cite{DBLP:conf/accv/XiaZKNZW18} and traffic accident anticipation on DADA-2000 \cite{DBLP:journals/tits/FangYQXY22}, CCD \cite{bao2020uncertainty}, and A3D \cite{DBLP:conf/iros/YaoXWCA19} datasets. 
\end{itemize}

The remainder of this work is as follows. Sec. \ref{relatedwork} briefly reviews the related works. Sec. \ref{approach1} and Sec. \ref{causeTAA} respectively describe the traditional supervised TAA formulation and the causal formulation for our equivariant TAA. The methodology is described in Sec. \ref{method}. Sec. \ref{exp} presents the experiments and the conclusion is given in Sec. \ref{con}.

\section{Related Work}
\label{relatedwork}
\subsection{Traffic Accident Anticipation}

Traffic accidents are a kind of scene anomaly~\cite{DBLP:journals/tcsv/LuCZZ23} that require a clear definition of the occurrence of a collision. Traditional research pipelines in road accident anticipation primarily focus on trajectory or frame prediction models~\cite{DBLP:journals/tmm/KimK21,DBLP:journals/tmm/00020WM023}, following a process involving detection, tracking, trajectory prediction, and accident or collision determination. Given the safety-critical nature of accidents, synthetic scenario generation also gathers significant interest in this field~\cite{aliakbarian2019viena,DBLP:conf/itsc/XiaoGYDW22}. Recently, TAA has entered the deep learning era, and the following two aspects are being featured in most formulations. 

\textbf{Object Feature Interaction in TAA}: Some works in TAA explore the interactive relationships within participant groups to determine the temporal consistency and movement risk~\cite{DBLP:conf/wacv/MallaCDCL23,DBLP:journals/tits/ChiaKGJ22}. Chan \emph{et al.}~\cite{DBLP:conf/accv/ChanCXS16} contribute the first accident anticipation model using Dynamic-Spatial-Attention (DSA) and Recurrent Neural Network (RNN) to correlate object interactions in driving videos. Inspired by the DSA module, Karim \emph{et al.}~\cite{karim2022dynamic} introduce a Dynamic Spatial-Temporal Attention (DSTA) network for traffic accident anticipation. Based on DSA, they model Dynamic Temporal Attention (DTA) to weigh the hidden states of past frames or object regions \cite{karim2023attention}. Furthermore, they propose a Spatial-Temporal Relational Learning (STRL) model~\cite{karim2022dynamic} 
 with Gated Recurrent Unit (GRU), which is inspired by the work~\cite{bao2020uncertainty} using the Graph Convolutional Recurrent Network (GCRN) ~\cite{seo2018structured} to model STRL. Similarly, Wang~\emph{et al.} \cite{wang2023gsc} also employs the Graph Convolutional Network (GCN) to model object feature interactions in driving scenes. They focus on learning the spatial-temporal continuity for edge weight updating in accident anticipation. Recently, DeepAccident~\cite{wang2023deepaccident} investigated accident anticipation in vehicle-to-vehicle scenarios focusing on trajectory collision. Through these deep learning strategies, TAA research aims to improve early predictions of potential collisions and provide valuable insights for safe decision-making in advanced driving systems.

\textbf{Frame Feature Consistency in TAA:}
Compared with modeling object-centric interactions in TAA, frame feature-consistency formulations relax the demand for accurate object detection, which eliminates possible detection errors in severe environments. Driver attention plays a crucial role in TAA since it reflects the driving task indirectly by indicating where drivers want to go or what they focus on. Frame feature consistency explores the hidden state consistency within the encoded frame features and pursues a core semantic learning for accident anticipation. In this domain, some works take driver attention as an auxiliary clue to assist the causal part learning of accident videos because driver attention in TAA models can help to identify dangerous objects without necessarily relying on object detection. Previous works~\cite{DBLP:journals/tits/FangYQXY22} in this area have explored the relationship between accident prediction and driver attention, yielding promising results. These understandings have inspired further development of TAA models; for example, \emph{e.g.,}, Bao~\emph{et al.} \cite{DBLP:conf/iccv/Bao0K21} leverage the driver attention to assist the accident anticipation via a deep reinforced learning model. A recent work~\cite{CognitiveTAA} further explores driver attention and textual description to facilitate TAA to form a cognitive TAA model. 

The aforementioned TAA works have shown promising performance, but they all rely on the well-annotated dataset, which limits their practical generalization in different driving scenes. In this work, we explore the video diffusion model for flexible data training without annotations.

\subsection{Video Diffusion Models}
Most video diffusion models are inspired by the Denoising Diffusion Probabilistic Models (DDPM) model~\cite{ho2020denoising}, which has a forward Markov process and a reverse process. \emph{Forward process} corrupts the original sample $x_0$ by gradually introducing noise to $x_k$ at $k$ step until the sample becomes completely random noise, and the \emph{reverse process} uses a series of Markov chains to remove the predicted Gaussian noise and recover the data progressively. With the diffusion operation, some label-efficient vision learning works~\cite{liao2022unsupervised} can be achieved by generating the pseudo-labels. Recently, video diffusion models have undergone the following two steps.

\textbf{Condition Control in Diffusion:} Based on DDPM, many conditional diffusion models emerge endlessly. The work~\cite{DBLP:conf/nips/DhariwalN21} trains a classifier to control labels with the diffusion guidance of the gradient $\nabla_{x_k}\log p_\phi(y|x_k, k)$ towards any category label $y$. Visual conditions~\cite{DBLP:conf/iccv/ChoiKJGY21,DBLP:conf/cvpr/LugmayrDRYTG22} add images in the sampling process and guide the reconstructed images to be similar to the reference images. Similarly, Kirstain \emph{et al.}~\cite{DBLP:journals/corr/abs-2303-01000} advocate the diffusion of various text-to-image scenarios that may benefit from visual cues, and some influencing works, \emph{e.g.}, DALL-E2~\cite{DBLP:journals/corr/abs-2204-06125} and Stable Diffusion~\cite{DBLP:conf/cvpr/RombachBLEO22}, have set an impressive milestone. Moreover, ControlNet~\cite{DBLP:journals/corr/abs-2302-05543} presents general support for additional input conditions. However, the above tasks are done in a single-flow framework. Recently, Versatile Diffusion~\cite{DBLP:journals/corr/abs-2211-08332} expands the single-flow diffusion into a multi-task multimodal network that handles multiple flows of text-to-image, image-to-text, and their variations in a unified model.

\textbf{Text-to-Video Generation}: Text-to-video generation is more challenging than image generation \cite{DBLP:journals/tmm/WangHZHW23} because it requires consistency in structure and spatiotemporal context. The pioneering work~\cite{DBLP:journals/corr/abs-2204-03458} presents the first results on video generation using diffusion models, which expands 2D U-net to 3D U-net. Then, Make-A-Video~\cite{DBLP:journals/corr/abs-2209-14792} leverages the help of many big diffusion models to synthesize videos conditioned on a given text prompt. Further, long video generation is explored by Yin \emph{et al.}~\cite{DBLP:journals/corr/abs-2303-12346}. However, most diffusion models require iterative computation which is costly for both training and inference. Stable diffusion models~\cite{DBLP:conf/cvpr/RombachBLEO22}, on the other hand, model the diffusion processes on the latent space, which can greatly reduce the computational complexity and generate images that can retain image diversity with high-quality visual semantics, which inspires the subsequent video diffusion works~\cite{DBLP:journals/corr/abs-2303-12688,DBLP:journals/corr/abs-2303-04761}.

With the video diffusion model conditioned by text prompts, we can generate diverse and large-scale accident or accident-free videos, and the generated videos can be treated as pseudo-positive or negative samples for the equivariant accident anticipation model.
\subsection{Causal Inference in Video/Image Understanding}
Causal inference recently is an attractive theory in image/video understanding for improving the model explainability and finding the important correlation between the causal variables to certain tasks, such as Video Question Answering (VQA)~\cite{DBLP:conf/emnlp/KoLKRK23,DBLP:journals/pami/ChenZNZX23,DBLP:conf/mm/Wei0YLL23,DBLP:conf/cvpr/ZangWPL23}, image recognition~\cite{DBLP:journals/tmm/QinZHLS23}, video anomaly detection~\cite{DBLP:conf/mm/0246XZ0WLJFLS23}, and object grounding~\cite{DBLP:journals/pami/WangGX23}, image captioning~\cite{DBLP:journals/pami/YangZC23}, \emph{etc.}
Under the Structural Causal Model (SCM)~\cite{glymour2016causal}, most video understanding works involved with causal inference theory aims to overcome the linguistic bias or visual content bias, through counterfactually masking the critical objects or words~\cite{DBLP:journals/pami/ChenZNZX23}, or object/video level grounding~\cite{DBLP:journals/pami/WangGX23,DBLP:conf/cvpr/ZangWPL23}, \emph{i.e.}, cutting off the backdoor path from the data domain to the videos/images and specific sentences by causal intervention. Then, the causal object or scenes are learned by contrastive learning for the complementary samples (\emph{i.e.}, the original samples, and the intervened ones). However, these formulations still rely on the pre-annotated labels.
As for this work, the EQ-TAA prefers a free-label accident anticipation,  where the causal intervention is achieved through the Attentive Video Diffusion (AVD) to change the normal video clips to accident clips driven by accident text prompts.

\section{Previous Supervised TAA}
\label{approach1}
Here, we briefly recap the supervised TAA prototype \cite{DBLP:conf/iccv/Bao0K21,wang2023gsc}. Given the video clip, TAA aims to maximize the Time-to-Accident (TTA) $\tau$, where $\tau$ = $\max\{0, t_{ai}-t\}$, $t$ denotes the first timestamp where the predicted accident probability $p_t$ is larger than a pre-defined threshold (\emph{e.g.}, 0.5 commonly), and $t_{ai}$ is the beginning time of the accident.  In the supervised TAA mode, the accident and accident-free samples are paired by video frames and the accident labels $y_t$ (1:accident, 0: accident-free). Consequently, given an observation video clip $V_{1:t}=\{I_1,..., I_{t}\}$, the objective of TAA is to predict the accident score $\hat{p}_t\in[0,1]$ and maximize $\tau$:
\begin{equation}
\hat{p}_t= f_y(V_{1:t}), 
\end{equation}
where $f_y$ is the combination of a video content encoder and the accident decoder. The video content encoder extracts the core semantics in the video frames up to time $t$ that may correlate with the occurrence of an accident at time $t_{ai}$. The accident decoder bridges the encoded video content with the accident occurrence. Notably, a video often has the consistent accident label $y_t$ for each frame. Therefore, the predicted $\hat{p}_t$ is treated as the predicted accident label $\hat{y}_t$ for a clear description.

\textbf{Learning}. To optimize the video content encoder and accident decoder, current works commonly adopt the scheme of empirical risk minimization (ERM) \cite{li2022invariant} between the predicted accident label $\hat{y}_t$ with the ground truth $y_t$:
\begin{equation}
\min \mathcal{L}_{\text{ERM}}(\hat{y}_t, y_t), \verb' 's.t., \max \tau.
\label{eq:1}
\end{equation}
Notably, the accident samples have an earliness penalty (maximizing $\tau$) of accident occurrence, commonly modeled by an exponential function \cite{DBLP:conf/cvpr/SuzukiKAS18}, while the accident-free samples usually formulate a Cross-Entropy (XE) to measure the label similarity. Essentially, previous supervised TAA works need ground-truth accident labels with laborious annotation, which is impractical in the open-traffic world.

\section{Causal Formulation for EQ-TAA}
\label{causeTAA}
We argue for research towards the Equivariant TAA work in this field with a more interpretable causal formulation. Specifically, just like the equivariant video representation learning \cite{DBLP:journals/tcsv/HuangHWYM22}, we do not need to annotate the accident label for each video frame. Instead, we create the accident label and advocate an equivariant accident label prediction after we intervene and create the causal part of each video. In light of this, we formulate the TAA with Structural Causal Model (SCM)~\cite{glymour2016causal} by checking the causal relation of four variables: input video $V$, background environment clip $B$, causal scene clip $C$, and the created accident label $y$.
  \begin{figure*}[!t]
  \centering
 \includegraphics[width=\hsize]{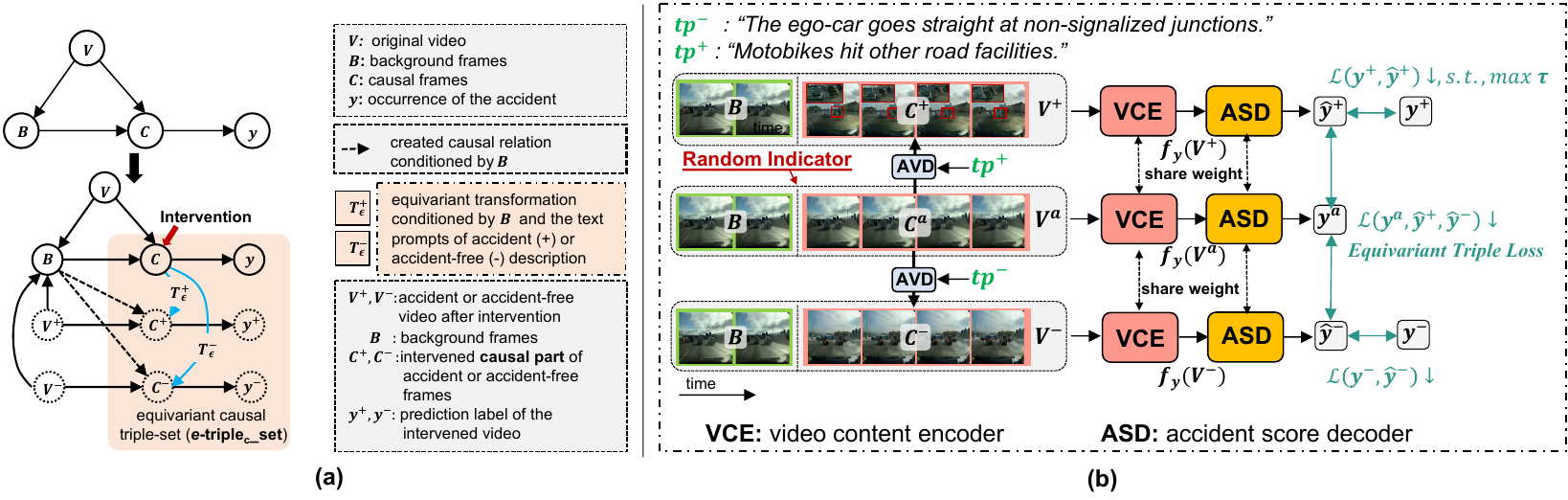}
  \caption{\small{The causal formulation of our Equivariant TAA, where (a) denotes the causal graph of EQ-TAA, and (b) is the training phase of EQ-TAA. ``\emph{tp+}" and ``\emph{tp-}"  denote the accident and accident-free text prompts, respectively. $\hat{y}^+$ and $\hat{y}^-$ are the predicted accident labels of the generated accident sample and the accident-free sample, and $y^a$ (accident-free commonly) is the accident label of the anchor video clip.}}
  \label{fig2}
  \vspace{-1.5em}
\end{figure*}

\textbf{Causal Formulation for TAA.}
We take a casual look at TAA (\emph{i.e.}, Fig. \ref{fig2}(a)).
\begin{itemize}
\item $B\leftarrow V \rightarrow C$: Each accident video $V$ can be divided into the set of background frames $\bf{\emph{B}}$ and the set of accident frames $\bf{\emph{C}}$, partitioned by a \emph{grounding indicator}.
\item $B\rightarrow C$: TAA implies the temporal correlation between the historical frames and causal frames. Therefore, $B$ is a backdoor causal relation to $C$ and $B\not\!\bot C$.
\item $C\rightarrow y$: The prediction of accident occurrence $y$ leverages the visual knowledge in causal part $C$.
\end{itemize}
The \emph{grounding indicator} aims to separate the background frames and causal frames. Instead of finding the grounding indicator by well-designed models \cite{li2022invariant,li2022acmmm}, we can take an arbitrary video frame in the normal videos as the starting frame of the accident/accident-free clip, and create the accident/accident-free frames flexibly by the attentive video diffusion model (Sec. \ref{evd}) guided by the accident/accident-free text prompts. Therefore, the grounding indicator here is denoted as \textbf{Random Indicator}. 
Assuming that we can create a pair of an accident clip (with $y^+$) and an accident-free clip (with $y^-$) for each anchor video clip $V^a$, we explore the rationalization for causal-equivariance of TAA.

\textbf{Rationalization.}  During training, the ground-truth grounding rationale $C$ is out of reach, while only the annotation-free input ($V^a$) is observed. Fortunately, we can collect a vast amount of accident-free videos and video-independent text descriptions of various accident types based on commonsense (\emph{e.g.}, positive text prompt: ``\emph{An ego-car hits a car}" and negative text prompt: ``\emph{A car moves straight on the road}", etc.). Such an ``absence" of accident labels motivates TAA to embrace text-guided video generation to create the rationale $C$. Besides, we also advocate a smooth temporal evolution of video content between $B$ and $C$. Hence, the text-guided video generation is also conditioned by some frames in the anchor clip $V^a$. Consequently,  we approach a pair of created accident and accident-free causal scenes, \emph{i.e.}, $C^+$ and $C^-$, and then generate prediction $y^+$ and ${y}^-$ via $y\leftarrow C$. To systematize this, an Attentive Video Diffusion (AVD) process $T_{\epsilon}$ is introduced based on the causal-equivariance principle to create $C^+$ and $C^-$, and expects an equivariant change in the response variable (\emph{i.e.}, $y^+$ and ${y}^-$). On top of SCM, we present such notions as:
\begin{equation}
T^+_{\epsilon}(y^+) = f_{y^+}(T^+_{\epsilon}(C^+)), \verb' 'T^-_{\epsilon}(y^-) = f_{y^-}(T^-_{\epsilon}(C^-)).
\end{equation}

\textbf{Extending ERM}: 
With the flexible causal part and accident label creation, we are excited that we can extend ERM with an Equivariant Triple Loss (ETL) that can enforce the causal part learning, \emph{i.e.}:
\begin{equation}
\min \mathcal{L}_{\text{ERM}}(\hat{y}^+_t, y^+_t)+\mathcal{L}_{\text{ERM}}(\hat{y}^-_t, y^-_t)+\mathcal{L}_{\text{ETL}}(y^a_t,
\hat{y}^+_t,\hat{y}^-_t),
\label{eq:2}
\end{equation}
where $y^a_t$ is the label of the anchor sample which is set as negative (-) for the easily collected accident-free videos. Besides, we can guarantee the sample balance for positive and negative samples all the time.

\section{Methodology}
\label{method}
We model the method flow for Equivariant TAA (EQ-TAA) in Fig. \ref{fig2}(b), where the timestamp of the Random Indicator is denoted as the beginning time of the accident. It contains three main components: 1) The Video Interventer by \textbf{AVD} partitioned by a Random Indicator, which creates the causal parts of the accident or accident-free samples, 2) the Video Content Encoding (\textbf{VCE}) and 3) the Accident Score Decoding (\textbf{ASD}). With this modeling, EQ-TAA is fulfilled with enforced causal part learning by the equivariant triple loss.

\begin{figure*}[!t]
  \centering
 \includegraphics[width=\hsize]{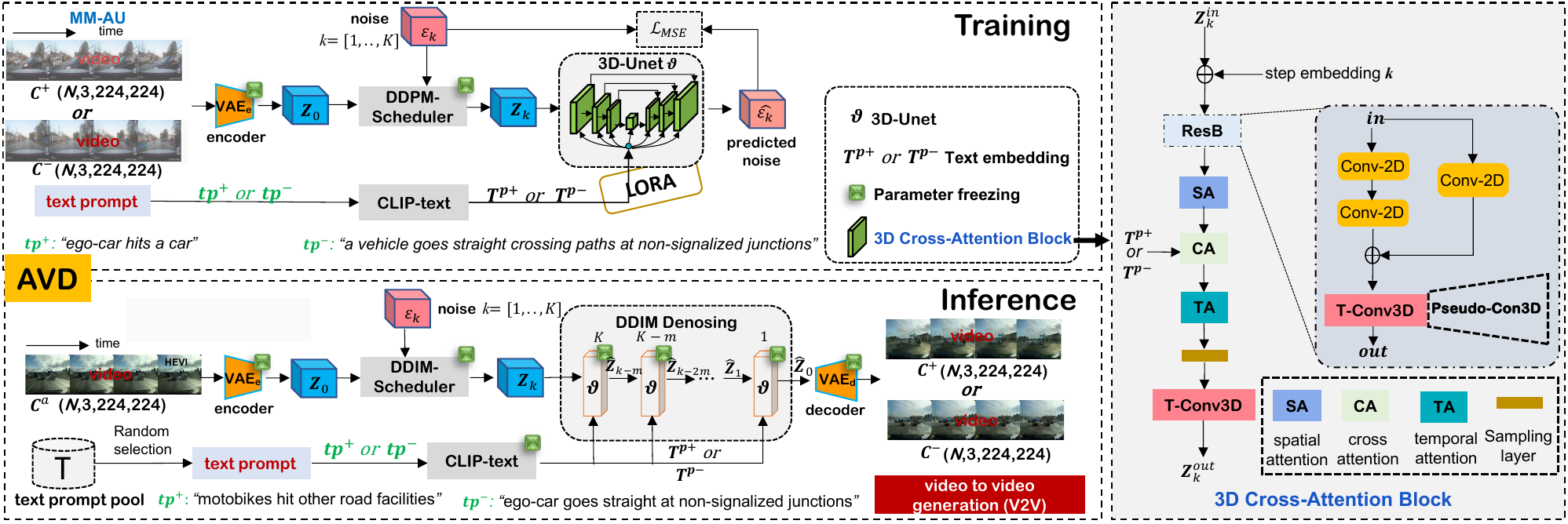}
  \caption{\small{The training and inference pipelines of AVD, where we utilize a Video-to-Video (V2V) generation framework for guaranteeing the temporal dependence among $N$ generated video frames. In the AVD training phase, we achieve the text-video alignment learning in the \textbf{Forward Diffusion} process by inputting accident ($C^+$) or accident-free ($C^-$) samples in the MM-AU dataset \cite{DBLP:journals/corr/abs-2212-09381}. In the AVD inference phase, we show the accident or accident-free video generation process through the \textbf{Reverse Diffusion} process (stepped by interval $m$, $m\gg 1$) inputted an anchor video clip ($C^a$) taken from arbitrary normal driving video datasets, such as HEV-I~\cite{yao2019egocentric}.}}
  \label{fig3}\vspace{-1.5em}
\end{figure*}
\subsection{Video Interventer by AVD}
\label{evd}
Here, we describe the Attentive Video Diffusion (AVD) model for causal part intervention (\emph{i.e.}, creation). Fig. \ref{fig3} shows the training and inference flowchart. The AVD model is inspired by the Stable diffusion \cite{DBLP:journals/corr/abs-2212-11565}, while we extend it into the video generation field with specially designed video encoding modules. 
 In the training phase, AVD is trained by a large-scale multimodality accident video dataset MM-AU \cite{DBLP:journals/corr/abs-2212-09381}, which contains 11,727 ego-view videos, 58 types of accident description, and 78 types of texts for describing normal traffic behaviors. These text prompts also constitute a text prompt pool $\mathcal{T}$. Each accident video is well-annotated for the accident and accident-free window. The text prompt pool $\mathcal{T}$ and the annotated accident videos are used for AVD Training. In the AVD Inference stage, we can randomly select a video clip in arbitrary normal ego-view driving videos, such as HEV-I~\cite{yao2019egocentric},  as the anchor clip $C^a$, and change its visual content to accident ($C^+$) or accident-free ($C^-$) conditioned by randomly selected accident ($tp^+$) or accident-free ($tp^-$) text prompts from $\mathcal{T}$. 

\subsubsection{AVD Training}
We use MM-AU \cite{DBLP:journals/corr/abs-2212-09381} with the well-annotated pairs of videos and text prompts ($tp^+$ for accident video and $tp^-$ for accident-free video) are adopted to train AVD for text-video content alignment learning. Leveraging the Latent Diffusion Model (LDM) \cite{DBLP:journals/corr/abs-2212-11565} for fine-tuning and text-to-video synthesis, our AVD also uses pre-trained LDM to map the videos into a latent space and model the diffusion process over the latent representation. 

AVD consists of three main components: 1) an autoencoder (VAE$_e$) that is used to map a video clip with $N$ frames into a latent representation $\textbf{Z}_0$; 2) a DDPM-scheduler that aims to forwardly add the noise $\varepsilon_k$ to $\textbf{Z}_0$ and generate noised latent representation $\textbf{Z}_k$ with $k$ steps for adding noise, and 3) a 3D U-net ${\bf{\vartheta}}$ which contains six 3D Cross-Attention Blocks (\textbf{3D-CAB}) (three encoder layers and three decoder layers) to estimate the noise $\hat{\varepsilon}_k$ and align the temporal text-video content correlation. AVD training aims to minimize the distance between $\hat{\varepsilon}_k$ and $\varepsilon_k$, where $k\in [1,..., K]$ denotes the diffusion step.

\textbf{LoRA Trick}: In the training phase, the text prompt is encoded by a CLIP-text model \cite{radford2021learning}, which is co-trained with the 3D U-net for self-style text-video content alignment learning. To preserve the knowledge in the latent representations and efficiently learn the alignment between videos and texts,  we use the Low-Rank Adaptation (LoRA) \cite{hu2021lora} fine-tuning trick. It takes the entire 3D U-net and CLIP-text model into the LoRA box which freezes the pre-trained LDM and CLIP-text model encoders and greatly improves the training efficiency for the fine-tuning diffusion models on our datasets. 

\subsubsection{AVD Inference}
Most modules in the AVD inference phase are similar to those of AVD training. Differently, the DDPM scheduler is switched into the Denosing Diffusion Implict Model (DDIM) scheduler \cite{DBLP:conf/iclr/SongME21} for a faster reverse diffusion process (\emph{i.e.}, changing the diffusion step interval 1 to $m$ and $m=50$ in this work), which leverages the trained 3D U-net and maintains better semantic consistency to the given video clips than DDPM. Similarly, the DDIM-scheduler also adds the noise $\varepsilon_k$ and estimates the noised latent representation $\textbf{Z}_k$. Differently, $\textbf{Z}_k$ is denoised to the clean representation $\hat{\textbf{Z}}_0$ by $K/m$ steps DDIM denoising process, where each step of denoising is fulfilled by the trained 3D U-net ${\bf{\vartheta}}$. $\hat{\textbf{Z}}_0$ is then decoded by VAE$_d$ to generate new $C^+$ or $C^-$ for the accident or accident-free clips. Notably, to preserve the temporal content dependency in the generated frames, we utilize a video-to-video generation pipeline, \emph{i.e.}, all $N$ frames of input video clip $C^a$ are fed into the AVD inference stage and have the same dimension as the generated ones. Here, the text prompt is randomly selected from the text prompt pool $\mathcal{T}$ and encoded by the same CLIP-text model trained in the AVD-training phase.

\subsubsection{3D-CAB}
It is worth noting that 3D-CAB is involved in the AVD training and inference phases. It is specially designed for spatial-temporal feature learning in our AVD model. A clear structure of 3D-CAB is illustrated in the right part of Fig. \ref{fig3}, which consists of a ResNet Block (ResB), a text-video Cross-Attention (CA) layer, a Spatial Attention (SA) layer, and a Temporal Attention (TA) layer. The core layer is the ResB layer which has a spatial-temporal feature encoding layer (T-Conv3D). In the original 3D U-net  \cite{DBLP:journals/corr/abs-2204-03458}, 3D convolution is adopted in T-Conv3D. However, it is time-consuming, and we instead introduce four layers of Pseudo-3D convolution \cite{DBLP:journals/corr/abs-2209-14792} to encode the spatial-temporal representation which replaces the $(3 \times 3 \times 3)$ 3D convolution with a $(1 \times 3 \times 3)$ 2D convolution and a $(3 \times 1 \times 1)$ 1D temporal convolution as:
\begin{equation}
\text{Conv-P3D}(\textbf{Z}_{in}) = \text{Conv-1D}(\text{Conv-2D}(\textbf{Z}_{in})^T)^T.
\end{equation}
Based on this strategy, we do not need to modify the 2D U-net shape and fulfill the 3D convolution effect.

\textbf{SA}:
Spatial Attention (SA) here aims to learn the self-attention between the spatial tokens in $\textbf{Z}_{in}^{k}$, where we denote the input of latent representation as  $\textbf{Z}_{in}^{k}$ and the output after self-attention as $\textbf{Z}_{out}^{k}$ in a consistent denotation.

Given $\textbf{Z}_{in}^{k}\in\mathbb{R}^{b\times N\times c \times H\times W}$ with the batch size $b$, $N$ frames, and $c$ channels of feature maps with size of $H\times W$, we reshape $\textbf{Z}_{{in}}^{k}$ in spatial attention as $\bar{\textbf{Z}}_{{in}}^{k}\in\mathbb{R}^{(b\times N)\times (H\times W) \times c}$, \emph{i.e.}, the token correlation in $b\times N$ frames is modeled over the  $(H\times W)\times c$ channels. Then, $\bar{\textbf{Z}}_{{in}}^{k}$ is encoded by a multi-head self-attention (\emph{e.g.}, $S$ heads) as:
\begin{equation}
 \textbf{Z}_{out}^{k}=\text{MHSA}^{(Spatial)}(\bar{\textbf{Z}}_{{in}}^{k})=\phi_{\textbf{W}_s}([h_1;h_2;\dots h_S]),
 \label{eq6}
\end{equation}
where $\phi_{\textbf{W}_s}$ is a linear transformation with parameters $\textbf{W}_s, s\in [1,S]$, and 
\begin{equation}
h_s=\text{SAtt}(\phi_{{\bf{W}}_{s{\bf{q}}}}(\textbf{Z}_{{in}}^{k}),\phi_{{\bf{W}}_{s{\bf{k}}}}(\textbf{Z}_{{in}}^{k}),\phi_{{\bf{W}}_{s{\bf{v}}}}(\textbf{Z}_{in}^{k})),
\end{equation}
where $\phi_{{\bf{W}}_{s{\bf{q}}}}(\textbf{Z}_{{in}}^{k})$,$\phi_{{\bf{W}}_{s{\bf{k}}}}(\textbf{Z}_{{in}}^{k})$, and $\phi_{{\bf{W}}_{s{\bf{v}}}}(\textbf{Z}_{in}^{k})$ denote the linear transformations of the query (\textbf{k}), key (\textbf{q}), and value (\textbf{v}) vectors of the $s$-$th$ self-attention (SAtt) head, respectively. 

SAtt is defined as:
\begin{equation}
\text{SAtt}(\textbf{Z}(\textbf{{q}}),\textbf{Z}(\textbf{k}),\textbf{Z}(\textbf{v}))=\delta(\textbf{Z}(\textbf{k})\textbf{Z}(\textbf{q})^T/\sqrt{d_{k}})\textbf{Z}(\textbf{v}),
\end{equation}
where $d_k$ is the head dimension of the query vector and $S=C/d_k$, and $\delta$ denotes the \emph{softmax} function.

\textbf{TA:} Temporal Attention (TA) is identical to SA while differs in token correlation which is modeled over the frame dimension.
Therefore, we reshape $\textbf{Z}_{in}^{k}\in\mathbb{R}^{b\times N\times c \times H\times W}$ into $\bar{\textbf{Z}}_{{in}}^{k}\in\mathbb{R}^{(b\times H\times W)\times N\times c}$, then the MHSA$^{(temporal)}(\bar{\textbf{Z}}_{{in}}^{k})$ operations are identical to Eq. \ref{eq6} with the same heads.

\begin{algorithm}[!t]
\caption{ \texttt{AVD} Algorithm}
\label{alg:EVD}
 \textbf{Input:} 
\\
\hspace{2em}- ${tp^+}$/${tp^-}$ : \text{text prompt for $C^+$ or $C^-$}.
\\
 {\bf Hyperparameters:} 
\\
\hspace{2em} - ${K}$: The number of diffusion steps;
 \\
\hspace{2em} - ${\beta}_{k}$:The variance across diffusion steps.
\\
 {\bf Forward  (\textcolor{teal}{AVD Training}):} 
\\

$\triangleright$ Get the text embedding of $tp^+$ or 
 $tp^-$.
\\
\hspace{2em} \textcolor{blue}{$\textbf{T}^{p+}$ or $\textbf{T}^{p-} = \text{CLIP}_{\text{text}}(tp^{+/-})$}
\\
$\triangleright$ Get the representation of the noise latent $\textbf{Z}_{t}$.

\hspace{2em}-$\textbf{Z}_0=\text{VAE}_e(C^{+/-})$;
\\
\hspace{2em}-$\textbf{Z}_{k} =$DDPM Schedule$ (\textbf{Z}_0,k,\varepsilon_{k})$
\\
\hspace{3em}$\textbf{Z}_{k} =\sqrt{\hat{\beta}_{k}} \textbf{Z}_0+\sqrt{1-\hat{\beta}_{k}}\varepsilon_{k}$,

where $\forall k = \{1,2,3,\dots,K\} \in {\text{Uniform}} $.
 $\hat{\beta}_{k}=\prod_{i=1}^{k}\alpha_{i};\alpha_{k}=1-\beta_{k}$; $\varepsilon_{k}\in{\mathcal{N}(0,I)}$
\\
$\triangleright$ \textbf{Output}:
\\
\hspace{2em}\textcolor{blue}{$\hat{\varepsilon_{k}}=\vartheta(\textbf{Z}_{k}, k, \textbf{T}^{p{+/-}})$}
\\
$\triangleright$ Optimization
\\
\hspace{2em} \textcolor{blue}{$\min\text{VAE}{\substack{\varepsilon \sim \mathcal{N}(0,I) \ k \sim \text{U}(1,K)}}\lVert \hat{\varepsilon_{k}} - \varepsilon_{k} \rVert_2^2$}
\\
 {\bf Reverse (\textcolor{teal}{AVD Inference}):} 
\\
 \textbf{Input:} The anchor video clip $C^a$.\\
 
$\triangleright$ Get the representation feature of the accident text or accident-free prompt $\textbf{T}^{p+}$ or $\textbf{T}^{p-}$.
\\
\hspace{2em}$\textbf{T}^{p+} or \verb' '\textbf{T}^{p-}=\text{CLIP}_{text}
(tp^{+/-})$,
\\ 
$\triangleright$ Denoising the latent token by video-to-video generation.
\\
\hspace{2em}-$\textbf{Z}_0=\text{VAE}_e(C^{a})$,
\\
\hspace{2em}-$\textbf{Z}_{k}=$DDIM Schedule$ (\textbf{Z}_0,k,\varepsilon_{k})$,
\\
\hspace{3em}$\textbf{Z}_k = \sqrt{\alpha_k}\textbf{Z}_{0} + \sqrt{1-\alpha_k}\varepsilon_k$,
\\
\textbf{For} $k=K,K-m,K-2m,...,1$ \textbf{do} \\
\hspace{3em} $\varepsilon_k\leftarrow \vartheta(\hat{\textbf{Z}}_{k}, k, 
\textbf{T}^{p{+/-}})$,
$\hat{\textbf{Z}}_{k}=\textbf{Z}_{k}$\\
\hspace{3em} $\hat{\textbf{Z}}_{k-1}=\sqrt{\alpha_{k-1}} \frac{\hat{\textbf{Z}}_{k}-\sqrt{1-\alpha_{k}\varepsilon_k}}{\sqrt{\alpha_{k}}}+\sqrt{1-\alpha_{k-1}\varepsilon_k}$\\
 \textbf{Output:} $C^{+/-}=\text{VAE}_d(\hat{\textbf{Z}}_0)$\\
\end{algorithm}
\textbf{CA}:
Text-Video Cross-Attention (CA) models the cross-modal attention \cite{DBLP:journals/tmm/LiuMZLWZ22} in text tokens $\textbf{T}^p$ and video tokens $\textbf{Z}_{{in}}^{k}$. 
To be computable, $\textbf{T}^p\in \mathbb{R}^{(b \times N) \times l \times c_{text}}$ is followed by a linear layer to transform the feature channels $c_{text}$ to $c$ of $\textbf{Z}_{{in}}^{k}$. Here, $l$ denotes the text prompt length, which is set to $77$ in this work. The Cross-Attention (CA) is defined as: 
\begin{equation}
\text{CA}(\textbf{Z}(\textbf{q}),\textbf{T}^p(\textbf{k}),\textbf{T}^p(\textbf{v}))=\delta(\textbf{Z}(\textbf{k})(\textbf{T}^p(\textbf{q}))^T/\sqrt{d_{k}})\textbf{T}^p(\textbf{v}).
\end{equation}
Different from SA and TA, CA receives query (\textbf{q}) from $\textbf{Z}_{k}$, while key (\textbf{k}) and value (\textbf{v}) are from $\textbf{T}^p$.
The clear implementation of AVD is illustrated in \textbf{Algorithm. 1}, where the \textcolor{blue}{blue} texts denote the parts with fine-tuning.

\subsection{Equivariant TAA}
\label{eq-taa}
As shown in Fig. \ref{fig2}(b), after completing the AVD training phase, we can then freely generate a vast number of accident $V^+$ or accident-free $V^-$ video clips respectively after grafting $N$ generated frames. Given the triple set of ($V^a$, $V^+$, $V^-$), we have three branches to encode the video content in them and predict the accident probability $y^a_t$, $\hat{y}^+_t$, and $\hat{y}^-_t$, respectively. To learn the core semantics of object regions and sudden movement for accident anticipation, we consider the uncertainty issue because sudden scene changes caused by critical objects may have direct links to future accidents. 
\begin{figure}[!t]
\centering
\includegraphics[width=\hsize]{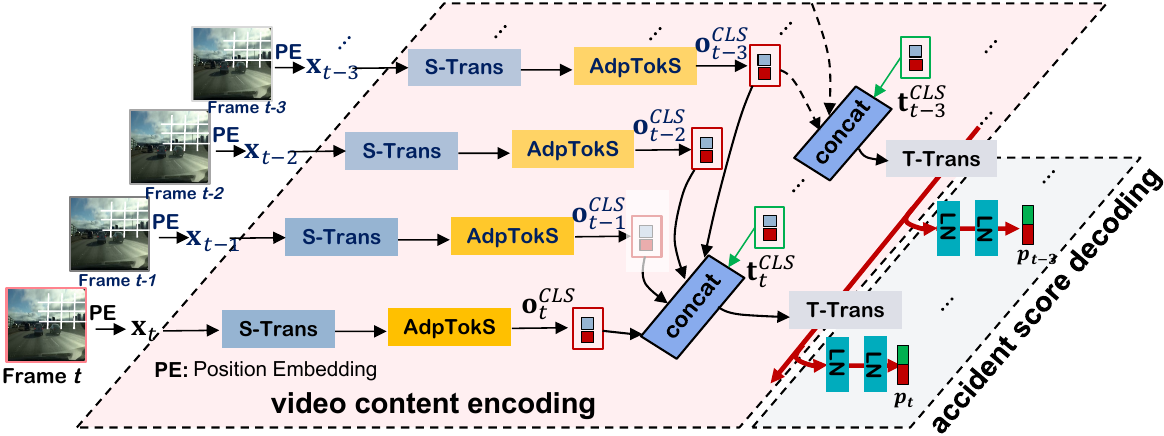}
\caption{\small{The video content encoding and accident score decoding modules, where \textbf{S-Trans} and \textbf{T-Trans} mean the spatial transformer and temporal transformer, and \textbf{AdpTokS} denotes an adaptive token sampling module for scene uncertainty consideration.}}
\label{fig4}
\end{figure}

\subsubsection{Uncertainty-aware Video Content Encoding (VCE)}
Fig. \ref{fig4} illustrates the flowchart of VCE, which contains a Spatial Transformer (S-Trans) block, an Adaptive Token Sampling (AdpTokS) block, and the Temporal Transformer (T-Trans) across frames of the token embeddings. Because it has the same structure as VCE in the triple set, we omit the ``a", ``+" or ``-" index and denote $V_{in}=\{v_{t}\}_{t=1}^n$ for the given input with $n$ frames (set as 5 in this work) in each encoding process.

\textbf{S-Trans}: We first encode each frame $v_t$ in $V_{in}$ to spatial tokens $\textbf{x}_t \in \mathbb{R}^{n \times Tok \times d}$($Tok=\frac{H}{P}\times\frac{W}{P}=196$ for 16 patch size $P$) with a linear projection $((3\times P\times P)\rightarrow d)$, where we set $d=192$. Then, we add the Position Embedding (PE)  $\textbf{x}_{PE}\in\mathbb{R}^{b \times n\times (Tok+1)\times d}$ and concatenate spatial cls-token $\textbf{x}_{cls}\in\mathbb{R}^{ b\times n \times 1\times d}$ to get new $\textbf{x}_t \in\mathbb{R}^{b \times n\times (Tok+1) \times d}$. Then we rearrange $\textbf{x}_t$ to $\hat{\textbf{x}}_t \in\mathbb{R}^{(b \times n) \times (Tok+1) \times d}$ for recurrently inputting to S-Trans block ($H_s$-layers). The operation of S-Trans is the same as the Spatial Attention (SA), \emph{i.e.}, 
\begin{equation}
\textbf{x}_t^o=\text{S-Trans}(\textbf{x}_t)=\text{MHSA}^{(spatial)}_{H_s}(\textbf{x}_t).
\end{equation}

\textbf{AdpTokS}: AdpTokS aims to select the important tokens with core object region semantics representation, which is formulated to consider the uncertainty of token learning. The token selection is based on the token attention ranking steps. Given $\textbf{x}_t^o$, it is filtered through a linear normalization. Then, the attention ranking leverages the self-attention of queries (\textbf{q}) and keys (\textbf{k}) scaled by $d_{0}$ at each attention head as:
\begin{equation}
\textbf{A}_t=\delta(\textbf{x}_t^o(\textbf{q})\textbf{x}_t^o(\textbf{k})^T/\sqrt{d_0}).
\end{equation}
The row summation of attention matrix $\textbf{A}_t\in\mathbb{R}^{(Tok+1)\times(Tok+1)}$ equals $1$, which represents the contribution of all input tokens to the output token. With $\textbf{A}_t$, the significance score of the $i^{th}$ token in each attention head at time $t$ can be calculated by:
\begin{equation}
a^{i}_t=\frac{\textbf{A}_t^{1,i} \times ||\textbf{x}_t^o(\textbf{v}_{i}) ||}{\sum_{i=2}^{Tok}\textbf{A}^{1,i}_t \times ||\textbf{x}_t^o(\textbf{v}_{i}) ||},
\end{equation}
where the token weight summation omits the ``cls-token" in the first column of $\textbf{x}_t^o$. The re-sampling of a token is based on its significance score summation over 8 heads, \emph{i.e.}, $e_i =\sum_{j=1}^{8} a^{j,i}_t$, where $j$ is the head index which is fixed as 8 and differs from the dynamic head index $s$ in Eq. \ref{eq6}.

Here, we do not adopt the cumulative token significance summation in \cite{fayyaz2021ats} for a convenient token sampling. To be adaptive for scene learning, similar to \cite{fayyaz2021ats}, we do not discard the tokens with lower $e_i$, while we only prune the attention weight $\textbf{A}_t$ to $\textbf{A}'_t$ with $Tok' (Tok'<Tok)$ tokens. Then, the significant token representation $\textbf{o}_t$ is obtained by $\textbf{A}'_t \textbf{x}_t^o(\textbf{v})$. In this work, we have 3 layers of AdpTokS to recurrently enforce the important token selection.

\textbf{T-Trans}: To model temporal information, we extract the ``cls-token" for each time $t$ and concatenate with the time cls-token $\textbf{t}^{CLS}_{t}$ to obtain the concatenated temporal video content representation $\textbf{O}_t$. $\textbf{O}_t$ is then fed into T-Trans ($H_t$-layers) for learning temporal relationships. The operation of T-Trans is the same as the $\text{MHSA}^{(temporal)}$. Finally, we use the max-pooling operation to obtain the salient token feature $\hat{\textbf{o}}^\kappa_t$ at time $t$ as:
\begin{equation}
\begin{aligned}
\textbf{O}_t=\text{Concat}(\textbf{o}^{CLS}_t,\textbf{o}^{CLS}_{t-1},...,\textbf{o}^{CLS}_{t-n},\textbf{t}^{CLS}_{t}),\\
\hat{\textbf{o}}_t=\text{T-Trans}(\textbf{O}_t)=\text{MHSA}^{(temporal)}_{H_t}(\textbf{O}_t),\verb' '\\
\hat{\textbf{o}}^{\kappa}_t=\text{maxpooling}(\hat{\textbf{o}}_t).\verb'              '
\end{aligned}
\end{equation}
After the video content encoding, $\hat{\textbf{o}}^{\kappa}_t$ is decoded to obtain the accident score at time $t$. 
\subsubsection{Accident Score Decoding (ASD)}
The accident score $\hat{p}_t$ is computed by decoding $\hat{\textbf{o}}^{\kappa}_t$ with two fully connected layers ($fc$) with the parameters of $\theta_1$ and $\theta_2$ and followed a \text{softmax} function, defined as:
\begin{equation}
\hat{p}_t=\delta(fc(fc(\hat{\textbf{o}}^{\kappa}_t;\theta_1);\theta_2)).
\end{equation}
\textbf{Notably}: To be consistent with Eq. 1, the accident score $\hat{p}_t \in [0,1]$ can be denoted as the predicted accident label $\hat{y}_t$.

\subsection{Optimization}
We re-formulate Eq. 4 for video content learning in EQ-TAA. 

\textbf{ERM Loss}: The ERM loss reflects the penalty for the prediction similarity of an accident or accident-free labels at each time. In particular, the accident video clip sample pursues an earliness of the accident. Therefore, the ERM loss is defined as a Cross-Entropy loss (XE) for accident-free samples and an exponential function loss \cite{DBLP:conf/cvpr/SuzukiKAS18} for accident samples:
\begin{equation}\small
\centering
\begin{aligned}
\mathcal{L}_{\text{ERM}}=w\mathcal{L}_{\text{ERM}}(y^-,{\hat{y}^{-}})+\mathcal{L}_{\text{ERM}}(y^+,{\hat{y}^{+}})\\
=-[w\frac{1}{N_I}\sum_{t=1}^{N_I}(1-y^-)\text{log}(1-\hat{y}_t^-)\\+\frac{1}{N_I}\sum_{t=1}^{N_I}y^+e^{-\text{max}(0,\frac{\tau}{r})}\text{log}(\hat{y}_t^+)],
\end{aligned}
\end{equation}
where $w$ is the weight of negative samples, and we set $w$ to 0.7. $N_I$ is the video sample length, $\tau$ is the time interval between current time $t$ to the beginning accident time $t_{ai}$, and $r$ is the frame rate.

\textbf{Equivariant Triple Loss}: In addition to ERM loss, we are excited that with the AVD model, we can generate a sample pair (an accident sample and an accident-free sample) for the anchor sample. Different from previous contrastive loss \cite{DBLP:journals/corr/abs-2302-02316,DBLP:journals/corr/abs-2302-02327}, we design an Equivariant Triple Loss (ETL) to distinguish the feature representation between $\hat{\textbf{o}}^{\kappa_+}_t$ and $\hat{\textbf{o}}^{\kappa_-}_t$. Different from ERM loss, we enforce ETL loss in the causal frames:
\begin{equation} \small
\begin{aligned} 
\mathcal{L}_{\text{ETL}} = \frac{1}{|t_{co}-t_{ai}|} \sum_{t=t_{ai}}^{t_{co}} [|| \hat{\textbf{o}}^{\kappa_a}_t, \hat{\textbf{o}}^{\kappa_-}_t ||^2 - || \hat{\textbf{o}}^{\kappa_a}_t,\hat{\textbf{o}}^{\kappa_+}_t ||^2 + \xi],
\end{aligned} 
\end{equation}
where $t_{co}$ is the collision time of an accident, and the parameter $\xi$ is used to ensure the discriminative difference between $\hat{\textbf{o}}^{\kappa_-}_t$ and $\hat{\textbf{o}}^{\kappa_+}_t$ with a pre-defined threshold (set as 0.5 commonly). $\hat{\textbf{o}}^{\kappa_a}_t$ denotes the token feature of anchor sample in each triple set. Thus, the optimization is achieved by minimizing:
\begin{equation}
\mathcal{L}={\mathcal{L}_{\text{ERM}}}_{[1,N_I]}+\lambda {\mathcal{L}_{\text{ETL}}}_{[t_{ai},t_{co}]},
\label{eq17}
\end{equation}
where $\lambda$ is a parameter for balancing the weight of two kinds of loss. We find that $\lambda$ is important with experimentally setting.
\section{Experiments}
\label{exp}

\subsection{Datasets}
We extensively evaluate our Attentive Video Diffusion (AVD) model and the Equivariant Traffic Accident Anticipation (EQ-TAA) to verify the performance. 
\subsubsection{For AVD}
As aforementioned, the AVD model is trained on a large-scale multimodality accident dataset (\textbf{MM-AU} \cite{DBLP:journals/corr/abs-2212-09381}) in driving scenes, where each video in MM-AU annotates the accident window [$t_{ai}$, $t_{ae}$]. To construct the text-video alignment pairs for training AVD, from MM-AU, we randomly select 5,842 accident-free frame clips in [1,$t_{ai}$] and 7,419 accident frame clips belonging to [$t_{ai}$, $t_{ae}$], resulting in a total of 13,271 text-video clips. Each clip consists of  $N=22$ frames with a corresponding text prompt. 

After AVD training, we can generate any number of accident or accident-free video clips by the videos in various driving scenes. In this work, we take the \textbf{HEV-I} \cite{yao2019egocentric} and \textbf{BDD-A} \cite{DBLP:conf/accv/XiaZKNZW18} datasets as the accident-free database and utilize the trained AVD to conduct a random sampling of 1,800 triple clip sets (\textbf{e-triple$_c$-sets}) from them to train the following EQ-TAA \footnote{The code will be available at \url{https://github.com/JWFanggit/Diffusion-TAA.}}. \textbf{HEV-I} \cite{yao2019egocentric} contains 230 videos, and the length of each video ranges from 10 to 60 seconds. \textbf{BDD-A} \cite{DBLP:conf/accv/XiaZKNZW18} contains 1,232 videos (=3.5 hours) collected from critical but accident-free scenarios in different occasions and light conditions.

For the training EQ-TAA, each video clip consists of $N_I=$150 frames (30fps), where the \emph{Random Indicator} (\emph{i.e.}, pseudo $t_{ai}$) is randomly determined in the frame range of [110,128] of each video clip, which results in pairs of accident and accident-free video clips with 22 synthetic frames starting from the pseudo $t_{ai}$. Notably, the Random Indicator can also be determined by an arbitrary video frame in HEV-I or BDD-A first, and then the associated video clip can be determined. 
\subsubsection{For EQ-TAA}
Our method is tested on the \textbf{DADA-2000} \cite{DBLP:journals/tits/FangYQXY22}, \textbf{A3D} \cite{DBLP:conf/iros/YaoXWCA19}, and \textbf{CCD} \cite{DBLP:conf/wacv/MallaCDCL23} datasets. 

\textbf{DADA-2000} has 2,000 accident videos, where for a fair comparison, same to DRIVE \cite{DBLP:conf/iccv/Bao0K21}, 1000 ones are selected for training and testing, where we sample 1800 pairs of positive and negative clips from the training set, and all the samples in DADA-2000 have 150 frames (5 seconds).

\textbf{A3D} contains 1,500 dashcam videos. Owing to the different temporal locations for the traffic accident in different videos, we trimmed each A3D video to 150 frames (5 seconds). Similar to the GSC \cite{wang2023gsc} work, we select 30\% of the A3D dataset as the testing set. 

\textbf{CCD} contains 1,500 dashcam accident videos. Each video is trimmed into 50 frames with 10 fps. Each accident is placed in the last 2 seconds of each accident video. Similar to the setting of CCD, 900 accident videos are chosen for testing.

\begin{figure}[!t]
  \centering
 \includegraphics[width=\linewidth]{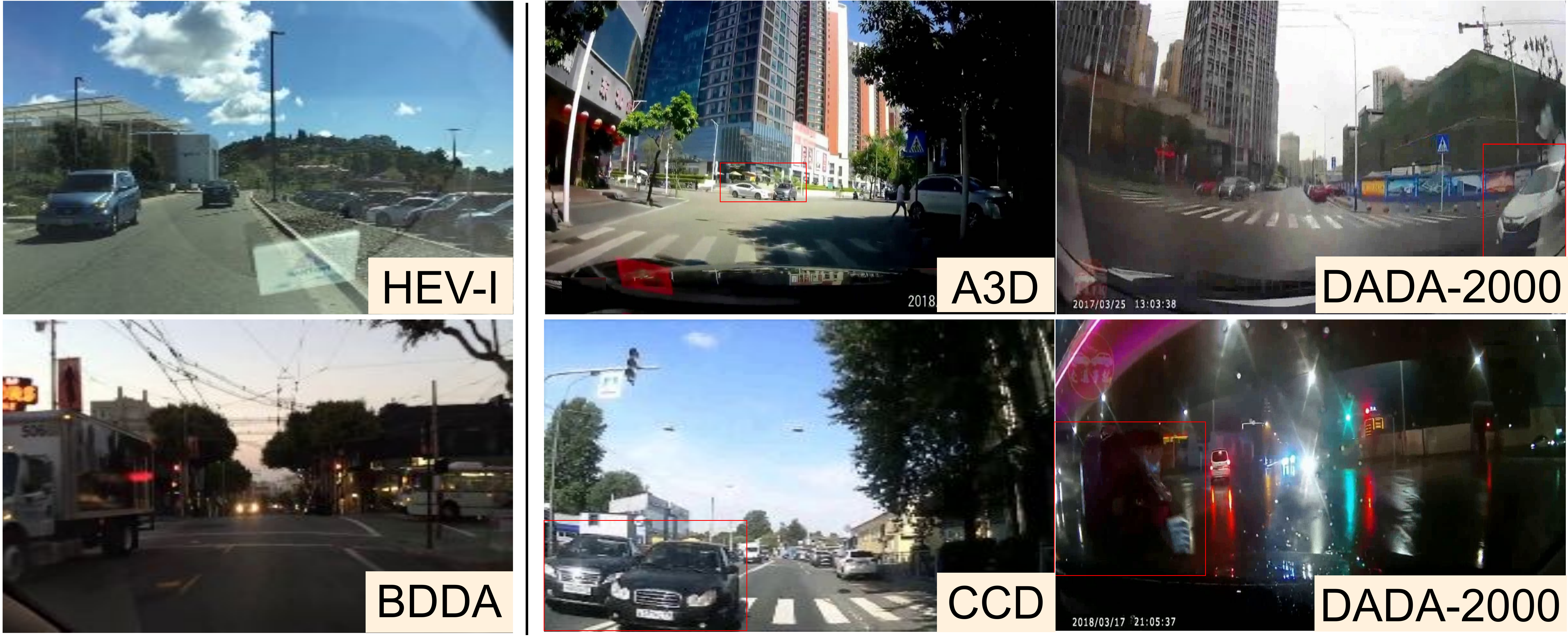}
  \caption{The frame snapshots of HEV-I \cite{yao2019egocentric}, BDD-A \cite{DBLP:conf/accv/XiaZKNZW18}, A3D \cite{DBLP:conf/iros/YaoXWCA19}, CCD \cite{DBLP:conf/wacv/MallaCDCL23}, and DADA-2000 \cite{DBLP:journals/tits/FangYQXY22}.}
  \label{fig5}
  \vspace{-1.5em}
\end{figure}

Notably, as shown in Fig. \ref{fig5}, most videos of A3D and CCD are collected in sunny daytime but DADA-2000 is more challenging with more diversified weather and light conditions. Therefore, we take the training set of DADA-2000 to train models, and A3D and CCD are only taken for testing.

\subsection{Metrics} 
This work adopts two groups of metrics for evaluation.

1) For AVD, we use the \emph{Fréchet Inception Distance (FID)}~\cite{heusel2017gans,yu2022efficient}, \emph{Fréchet Video Distance (FVD)}~\cite{unterthiner2018towards}, and the \emph{Inception Score (IS)}~\cite{DBLP:conf/nips/SalimansGZCRCC16} as a quality evaluation for the synthetic frames and video clips, respectively. FVD uses a pre-trained I3D Network \cite{carreira2017quo} to calculate the mean vector and covariance matrix of the real video clip $C^a$ (22 frames) and the generated $C^+$ or $C^-$. FID uses a pre-trained InceptionV3 Network \cite{szegedy2016rethinking} to calculate the mean vector and covariance matrix of each frame. IS also takes the pre-trained InceptionV3 Network to calculate the exponential value of Kullback–Leibler divergence (KLD) between the generated images and real images. In addition, IS also reflects the distribution diversity of the generated images.
Furthermore, we introduce the Contrastive Language-Image Pre-Training score (CLIP$_S$) to measure the alignment degree between text prompts and video frames. Larger CLIP$_S$ values indicate better coherent alignment. The CLIP$_S$ is defined as:
\begin{equation}
\text{CLIP}_S = \frac{1}{N}\sum_{i=1}^{N}\text{sim}(f_{\theta}(I_i),f_{\theta}(tp)),
\end{equation}
where $N$ represents the number of video frames, $\text{sim}(.,.)$ represents the cosine similarity function, $f_{\theta}$ denotes the feature extractor function defined by the CLIP model \cite{radford2021learning}. $I_i\in C^{+/-}$ is the $i^{th}$ video frame, and $tp$ specifies the text prompt.

2) For TAA, we employ the same metrics used in previous works \cite{bao2020uncertainty,DBLP:conf/iccv/Bao0K21}, \emph{i.e.}, Average Precision (AP) \cite{bao2020uncertainty}, and Time-to-Accident (TTA) \cite{DBLP:conf/iccv/Bao0K21}, which includes the variants of TTA$_{a}$ and mTTA. AP evaluates the average accuracy with different thresholds of precision and recall rate. TTA$_{a}$ is computed by setting the accident score threshold in each video as $a$ to evaluate the \emph{earliness} of a positive prediction, which is determined by the time interval between \emph{the first warning frame} (with the accident score larger than $a$) and the starting time of the accident ($t_{ai}$). mTTA denotes the mean TTA, which is computed by the Expectation of TTA$_{a}, a\in [0,1]$.  In this work, we also evaluate the methods based on video-level Area under ROC curve (AUC) \cite{DBLP:conf/iccv/Bao0K21} for DADA-2000. AUC is a metric for evaluating the performance of accident or accident-free determinations. These metrics pursue a larger value. 

In addition, TAA is a temporal forecasting problem and owns a prediction uncertainty issue. In this work, we further introduce a prediction Variance (\textbf{Var}) when the accident scores of frames are larger than a pre-defined threshold (set as 0.5 in this work), and stable anticipation (low Var values) is preferred for reducing false alarms in driving.

\subsection{Implementation Details}
\begin{figure}[!t]
  \centering
 \includegraphics[width=0.9\linewidth]{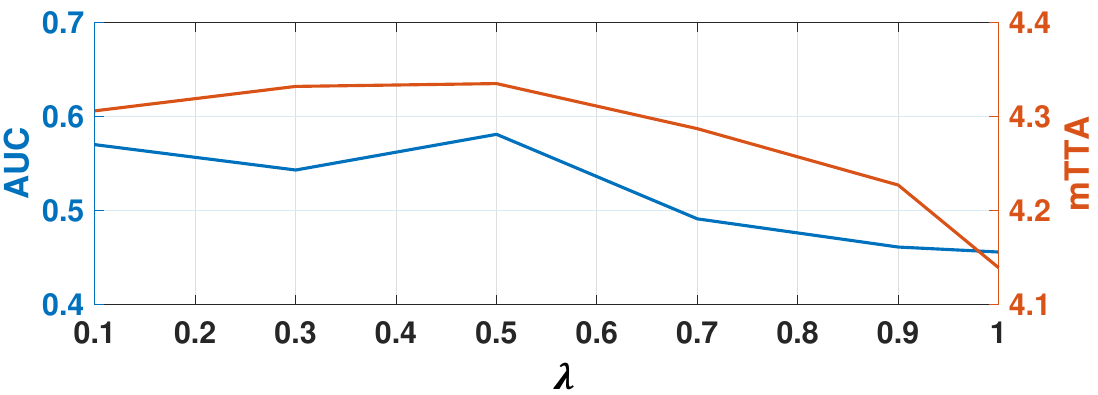}
  \caption{The AUC and mTTA values, \emph{w.r.t.,} $\lambda$, on the testing set of DADA-2000.}
  \label{fig6}
\end{figure}

In this work, the ETL loss in Eq. \ref{eq17} is the key component for the TAA equivariance of this work, where the parameter $\lambda$ acts as a trade-off between the ETL loss and ERM loss. Therefore, we experimentally check the role of $\lambda$ by setting it as [0.1,0.3,0.5,0.7,0.9]. Here, we evaluate the performance difference on the testing set of the DADA-2000 with the AUC and mTTA metrics. Fig. \ref{fig6} shows that $\lambda=0.5$ obtains the best values and is adopted for more consideration of ETL loss.

\textbf{Configuations for AVD:} For training AVD, we set the learning rate as $5e^{-6}$, and the batch size is set as 2. Xformers \cite{xFormers2022} is utilized to optimize hardware memory efficiency. The number of training steps $K$ is 8000. During the AVD inference, we generate videos with the same configuration, while the inference step interval $m$ is set as 50.

\textbf{Configuations for EQ-TAA:} The layers of S-Trans (\emph{H}$_s$) and T-Trans (\emph{H}$_t$) are set as 2 and 3, respectively. The dimensions of two fully connected layers ($fc$) are transformed as 192$\rightarrow$64$\rightarrow$2. The ETL loss parameter  $\xi$ is set to 0.5. We use the AdamW optimizer with a learning rate of $1e^{-4}$, $\beta_1$ of $0.9$, $\beta_2$ of $0.999$, and weight decay of $0.01$. Because of the large-scale samples, all experiments in EQ-TAA are trained for one epoch with a batch size of 1. 
  \begin{figure*}[!t]
\centering
\includegraphics[width=0.95\hsize]{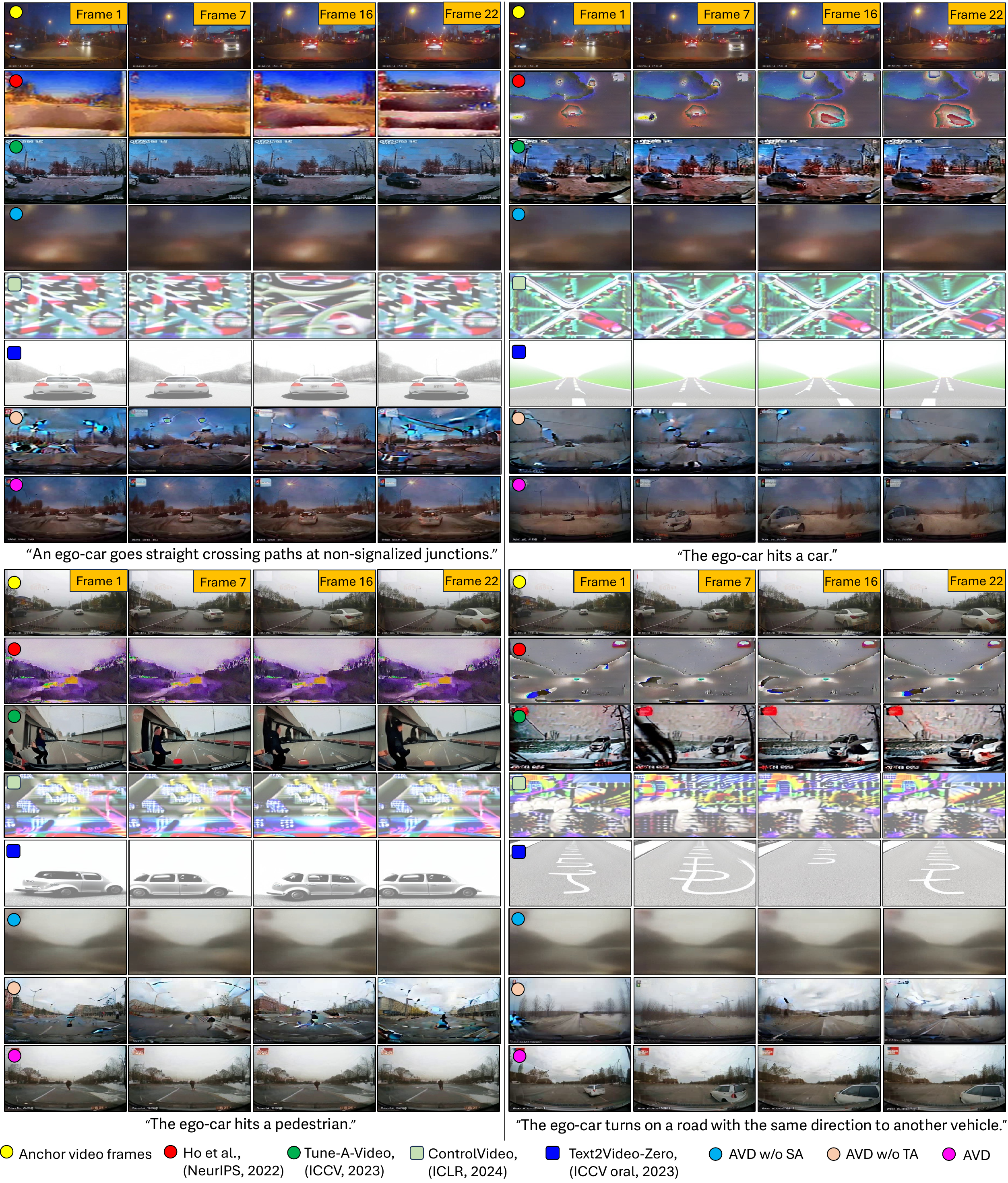}
\caption{The snapshots of different video diffusion models on the two videos of BDD-A dataset \cite{DBLP:conf/accv/XiaZKNZW18}, where the results in each block are obtained by the method of Ho \emph{et al.}  \cite{DBLP:journals/corr/abs-2204-03458}, Tune-A-Video \cite{DBLP:journals/corr/abs-2212-11565}, ControlVideo~\cite{ControlVideo2024}, Text2Video-Zero~\cite{DBLP:conf/iccv/KhachatryanMTHW23}, our AVD and its variants (Best viewed in color mode).}
\label{fig7}
\vspace{-1.5em}
\end{figure*}

  \begin{figure*}[!t]
\centering
\includegraphics[width=0.95\hsize]{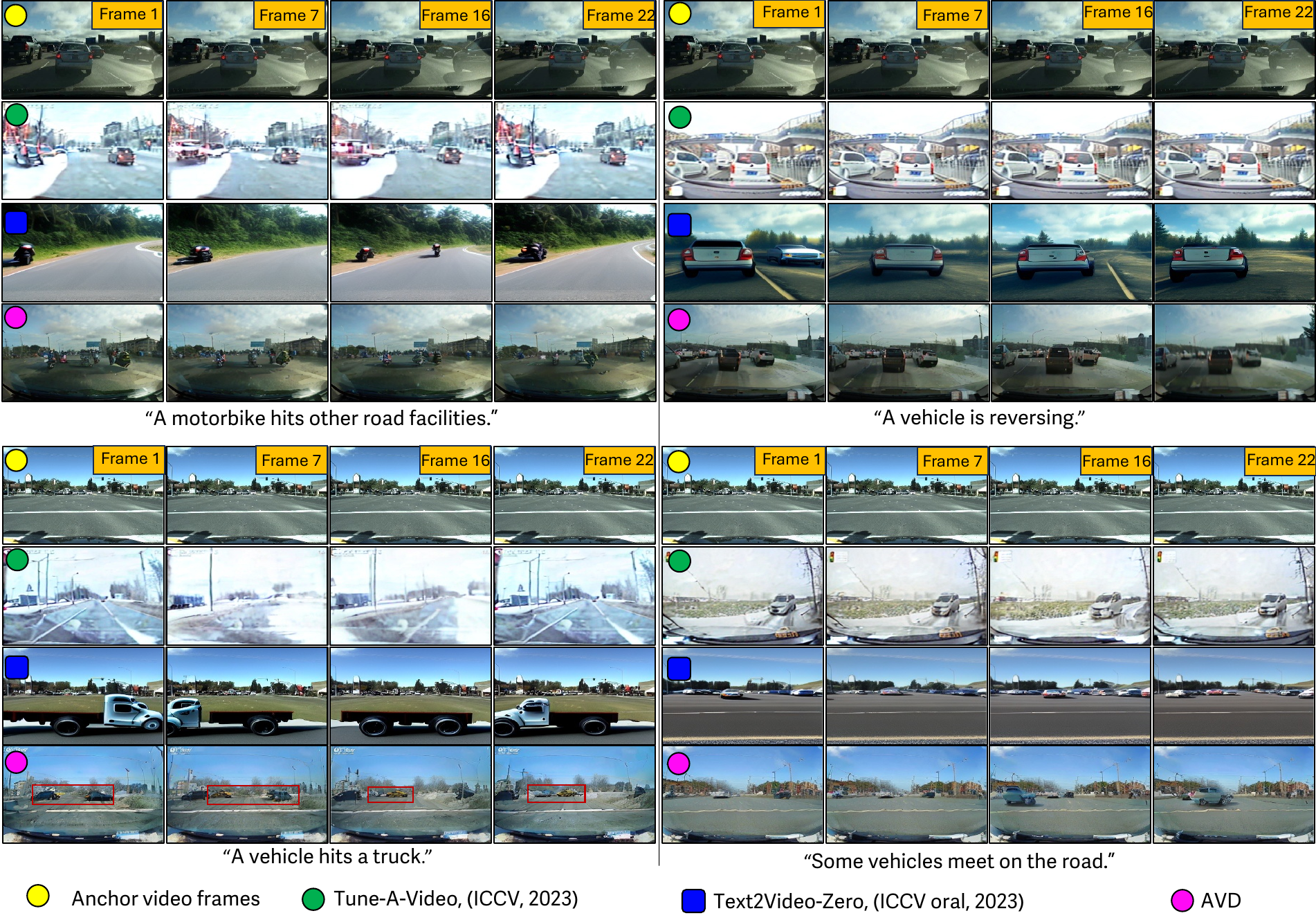}
\caption{The snapshots of different video diffusion models on the two videos of BDD-A dataset \cite{DBLP:conf/accv/XiaZKNZW18}, where the results in each block are obtained by the method of Tune-A-Video \cite{DBLP:journals/corr/abs-2212-11565}, Text2Video-Zero~\cite{DBLP:conf/iccv/KhachatryanMTHW23}, and our AVD (Best viewed in color mode).}
\label{fig8}
\end{figure*}

All experiments are performed on a platform with one Nvidia RTX 3090Ti GPU and 64G RAM.

\subsection{Evaluation for AVD}
In this work, we select four state-of-the-art video diffusion models including Tune-A-Video (\emph{abbrev.} TAV) \cite{DBLP:journals/corr/abs-2212-11565}, Ho et al. \cite{DBLP:journals/corr/abs-2204-03458}, ControlVideo (Cvideo)~\cite{ControlVideo2024}, and Text2Video-Zero (T2V-Zero)~\cite{DBLP:conf/iccv/KhachatryanMTHW23} to compare with our AVD. To be fair, we also take the anchor video frames as the condition for these diffusion models. TAV takes the DDIM inversion to generate video frames. For each video diffusion model, we generate 1,800 clips, and each contains 22 frames. Cvideo and T2V-Zero are the training-free video diffusion methods. In addition, we also evaluate the roles of different self-attention models in our AVD, which results in two versions of AVD, namely ``\emph{AVD w/o SA}" (without MHSA$^{(Spatial)}$ defined in Eq. \ref{eq6}) and ``\emph{AVD w/o TA}" (without MHSA$^{(temporal)}$). 
\begin{table}[!t]\small
  \centering
  \caption{Performance comparison for video diffusion models with 1800 generated clips sampled from \textbf{HEV-I} \cite{yao2019egocentric} and \textbf{BDDA} \cite{DBLP:conf/accv/XiaZKNZW18}.}
      \renewcommand{\arraystretch}{1.2}
 \setlength{\tabcolsep}{1.2mm}{
\begin{tabular}{l|c|c|c|c|c}
\toprule[0.8pt]
Method  & CLIP$_S$$\uparrow$&FID$\downarrow$&FVD$\downarrow$&IS$\uparrow$&FPS$\uparrow$\\
\hline\hline
Ho et al.~\cite{DBLP:journals/corr/abs-2204-03458} &21.77&2098.65&9202.25&3.75&1.0\\
TAV~\cite{DBLP:journals/corr/abs-2212-11565}&24.80&854.05&8810.35&3.02&2.7\\
CVideo~\cite{ControlVideo2024}&22.06&928.12&10551.49&4.11&0.5\\
T2V-Zero~\cite{DBLP:conf/iccv/KhachatryanMTHW23}&26.22&\textbf{585.73}&9842.69&\textbf{4.33}&1.1\\
\hline
AVD w/o TA  &28.95&806.88&9651.40&2.01&4.4\\
AVD w/o SA  &21.82&917.55&10648.71&3.13&4.4\\
\textbf{AVD} &\textbf{29.63}&673.35&\textbf{7924.75}&2.36&4.4\\
\toprule[0.8pt]
  \end{tabular}}
  \label{tab1}
  \end{table}
  
Table. \ref{tab1} presents the quantitative metric results, from which we can observe that our AVD can obtain significant improvement from generation quality (FVD), text-video alignment, and computation efficiency. 
We can see that the MHSA$^{(Spatial)}$ offers stronger supports to AVD model. We present some snapshots of generated video frames in Fig.  \ref{fig7}. It is clear that the MHSA$^{(Spatial)}$ can enforce the textual details of generated frames. Without MHSA$^{(Spatial)}$, the generated frames show an unclear appearance. Without MHSA$^{(temporal)}$, many artifacts appear. For instance,  after adding the accident text prompt ($tp^{+}$) of ``\emph{The ego-car hits a pedestrian}"),  The version of ``AVD w/o SA" is unresponsive for the crossing pedestrian. ``AVD w/o TA" cannot generate a vehicle when adding the accident-free text prompt ($tp^-$) ``\emph{The ego-car turns on a road with the same direction to another vehicle}.”

Notably, the IS values of our AVD are smaller than other models. However, in the visual comparison in Fig.~\ref{fig7}, our AVD is better at preserving the content style of the anchor video frames than the Cvideo, TAV, T2V-Zero, and Ho et al.'s approach. AVD also enforces a responsive change of the specific object reflected by the largest CLIP$_S$ value. Although the anchor frames are taken as the input condition, the methods of Ho \emph{et al.},  T2V-Zero, and Cvideo tend to output irrelevant frame content while TAV generates many artifacts. Although T2V-Zero obtains the best FID value, it has no relation to the anchor video frames and outputs irrelevant visual content, as shown in Fig.~\ref{fig7}. Therefore, the IS metric is easily influenced by the noisy video content.

To further check the visual comparison of different video diffusion models, we compare Tune-A-Video (TAV), and Text2Video-Zero (T2V-Zero) in Fig. \ref{fig8}. The results show that T2V-Zero generates the video frame with the highest quality but it does not preserve the anchor video frame style and the text prompts are not responsive in the generated frames. TAV again shows many artifacts and the generated video content has a large distance to the anchor video frames. 

\subsection{Overall Evaluation for EQ-TAA}
\subsubsection{Competitors}
Since this work contributes an Equivariant TAA (EQ-TAA) method, we compare its performance with several state-of-the-art supervised and Equivariant TAA (EQ-TAA) methods driven by Tune-A-Video (TAV) \cite{DBLP:journals/corr/abs-2212-11565}, Ho \emph{et al.} \cite{DBLP:journals/corr/abs-2204-03458}, ControlVideo (Cvideo)~\cite{ControlVideo2024}, and Text2Video-Zero (T2V-Zero)~\cite{DBLP:conf/iccv/KhachatryanMTHW23}. Based on the main formulation, we categorize the supervised approaches to \textbf{Object-centric TAAs (Ob.TAAs)}, and \textbf{Frame-centric TAAs (Fr. TAAs)}. Object-centric TAAs are as follows. To be fair, we try my best to balance the training samples in supervised approaches with real positive and negative video clips and our EQ-TAA
driven by generated video clips ($\sim$1800 pairs of negative and positive clips for training all methods).

\textbf{DSA} \cite{DBLP:conf/accv/ChanCXS16} is the first deep learning-based TAA work in dashcam videos, which model a Dynamic Spatial Attention (DSA) incorporating the Recurrent Neural Network (RNN) to correlate the spatial-temporal correlation of objects in frames.

\textbf{UncertaintyTA} \cite{bao2020uncertainty} models a graph-based spatial-temporal relation learning model, and considers the uncertainty estimation for the predicted accident scores.

\textbf{DSTA} \cite{DBLP:journals/tits/KarimLQY22} formulates a dynamic spatial-temporal attention model to infer the correlation of object interaction between video frames, and \textbf{GSC} \cite{wang2023gsc} models a spatio-temporal graph continuity to infer the temporal consistency of video frames.

In addition to object-centric TAAs, three frame-centric TAAs, \emph{i.e.}, \textbf{DRIVE} \cite{DBLP:conf/iccv/Bao0K21}, \textbf{XAI-AN} \cite{DBLP:journals/corr/abs-2108-00273}, and \textbf{Cog-TAA} \cite{CognitiveTAA}, are selected. Different from the object-centric approaches which need to pre-detect the object accurately, frame-centric TAAs formulate the accident anticipation by spatial-temporal frame feature encoding and accident score decoding frameworks. DRIVE \cite{DBLP:conf/iccv/Bao0K21} involves the driver fixation map specially to assist the core semantic region learning, and Cog-TAA \cite{CognitiveTAA} introduces the text description as input and takes the fixation attention to guide critical object learning for cognitive TAA.

\subsubsection{Result Analysis}
For the DADA-2000 dataset, the metrics of AP, AUC, TTA$_{0.5}$, and mTTA are all used, while the AUC metric is omitted for the CCD and A3D datasets because the testing samples in them are all positive accident videos. Because we adopt the video clips with synthetic video frames to train the model, we make an equivalent training data scale to the training sets of the A3D, CCD, and DADA-2000 datasets. We use the results reported in other works on A3D and CCD datasets for comparison. For the DADA-2000 dataset, the results of other methods are reported in DRIVE~\cite{DBLP:conf/iccv/Bao0K21} and Cog-TAA~\cite{CognitiveTAA} with the equivalent training samples. 
\begin{table}[!t]\small
\centering
\caption{The performance evaluation for different TAA Categories (TAA. Cat) in the testing set of the A3D \cite{DBLP:conf/iros/YaoXWCA19} and CCD \cite{DBLP:conf/wacv/MallaCDCL23} datasets, where the best value of each metric is marked in \textbf{bold} font.}
\renewcommand{\arraystretch}{1.2}
     \setlength{\tabcolsep}{0.4mm}{
\begin{tabular}{c|l|cc|cc}
\toprule[0.8pt]
    \multirow{2}[4]{*}{TAA. Cat.}&  \multirow{2}[4]{*}{Method} & \multicolumn{2}{c|}{A3D \cite{DBLP:conf/iros/YaoXWCA19}} & \multicolumn{2}{c}{CCD \cite{DBLP:conf/wacv/MallaCDCL23}}\\
\cmidrule{3-6}        &  &AP$\uparrow$ &mTTA$\uparrow$ & AP$\uparrow$ &mTTA$\uparrow$ \\
\hline
  \multirow{2}[4]{*}{}&UncertaintyTA \cite{bao2020uncertainty}&0.944&\textbf{4.920}&0.995&4.740\\
  \multirow{2}[4]{*}{Ob.TAAs}&DSA \cite{DBLP:conf/accv/ChanCXS16}&0.934&4.410&0.996&4.520\\
  \multirow{2}[4]{*}{}&DSTA \cite{DBLP:journals/tits/KarimLQY22}&-&-&0.996&\textbf{4.870}\\
  \multirow{2}[4]{*}{}&GSC \cite{wang2023gsc}&-&-&0.949&2.620\\
\hline
  \multirow{2}[4]{*}{Fr. TAAs}&XAI-AN \cite{DBLP:journals/corr/abs-2108-00273}&-&-&0.940&4.570\\
  \multirow{2}[4]{*}{}&DRIVE \cite{DBLP:conf/iccv/Bao0K21}&0.918&4.045&0.992&4.490\\
  \multirow{2}[4]{*}{}&Cog-TAA \cite{CognitiveTAA}&0.920&4.128&0.997&4.630\\
\hline
  \multirow{2}[4]{*}{}&Driven by Ho \emph{et al.} \cite{DBLP:journals/corr/abs-2204-03458}&0.907&3.865&0.985&4.108\\
  \multirow{2}[4]{*}{EQ-TAA}&Driven by TAV \cite{DBLP:journals/corr/abs-2212-11565}&0.905&4.005&0.986&4.301\\
    \multirow{2}[4]{*}{}&Driven by CVideo~\cite{ControlVideo2024}&0.903&3.626&\textbf{0.999}&2.814\\
        \multirow{2}[4]{*}{}&Driven by T2V-Zero~\cite{DBLP:conf/iccv/KhachatryanMTHW23}&0.906&3.536&\textbf{0.999}&2.685\\
  \multirow{2}[4]{*}{}&Driven by \textbf{AVD}&\textbf{0.953}&4.321&0.989&4.407\\
\toprule[0.8pt]
  \end{tabular}}
  \label{tab2}
  \end{table}

  \begin{table}[!t]\small
\centering
\caption{The performance evaluation for different TAA Categories (TAA. Cat) in the testing set of the DADA-2000 \cite{DBLP:journals/tits/FangYQXY22} dataset, where the best value is marked in \textbf{bold} font.}
\renewcommand{\arraystretch}{1.2}
     \setlength{\tabcolsep}{0.3mm}{
\begin{tabular}{c|l|cccc}
\toprule[0.8pt]
   \multirow{2}[4]{*}{TAA. Cat.}&\multirow{2}[4]{*}{Method} & \multicolumn{4}{c}{DADA-2000 \cite{DBLP:journals/tits/FangYQXY22}}\\
\cmidrule{3-6}    &      & AP$\uparrow$ &AUC$\uparrow$  & TTA$_{0.5}\uparrow$ & mTTA$\uparrow$\\
\hline
  \multirow{2}[4]{*}{Ob.TAAs}& UncertaintyTA \cite{bao2020uncertainty}&-&0.600&3.849&-\\
  \multirow{2}[4]{*}{}&AdaLEA \cite{DBLP:conf/cvpr/SuzukiKAS18}&-&0.550&3.890&-\\
  \multirow{2}[4]{*}{}&DSA \cite{DBLP:conf/accv/ChanCXS16}&-&0.470&3.095&-\\
\hline
  \multirow{2}[4]{*}{Fr. TAAs}&DRIVE \cite{DBLP:conf/iccv/Bao0K21}&0.690&0.727&3.657&4.259\\
  \multirow{2}[4]{*}{}&Cog-TAA \cite{CognitiveTAA}&0.744&\textbf{0.833}&\textbf{4.020}&\textbf{4.497}\\
\hline
  \multirow{2}[4]{*}{}&Driven by Ho \emph{et al.} \cite{DBLP:journals/corr/abs-2204-03458}&0.722&0.500&3.724&4.159\\
  \multirow{2}[4]{*}{EQ-TAAs}&Driven by TAV \cite{DBLP:journals/corr/abs-2212-11565}&\textbf{0.752}&0.532&3.668&4.286\\
      \multirow{2}[4]{*}{}&Driven by CVideo~\cite{ControlVideo2024}&0.716&0.448&1.462&3.546\\
        \multirow{2}[4]{*}{}&Driven by T2V-Zero~\cite{DBLP:conf/iccv/KhachatryanMTHW23}&0.698&0.421&1.023&3.481\\
  \multirow{2}[4]{*}{}&Driven by \textbf{AVD} &0.725&0.580&3.894&4.258\\
\toprule[0.8pt]
  \end{tabular}}
  \label{tab3}
  \end{table}

As shown in Table. \ref{tab2} and Table. \ref{tab3}, compared with the supervised ones, the equivariant methods still have a distance towards achieving the best performance obtained by Cog-TAA, especially for the AUC metric as shown in Table. \ref{tab3}. It indicates that because of the synthetic nature, the generated accident videos and accident-free videos still need to be enhanced from their inter-class difference with more well-designed text-video coherent alignment models. 

\textbf{EQ-TAA (AVD) vs. other EQ-TAAs}: We compare the roles of different video diffusion models for TAA here. Actually, for the TAA task, the label of each frame is influenced by the content difference between the historical frame observation and the accident frames. From Fig. \ref{fig7}, we can see that although TAV~\cite{DBLP:journals/corr/abs-2212-11565} generates many artifacts, it shows a significant difference between the generated ones and the anchor video frames. Therefore, it obtains the best AP value and mTTA value for the DADA-2000 dataset, as shown in Table. \ref{tab3}. Besides, the generated video frames need to reflect the causal objects well. Therefore, although CVideo~\cite{ControlVideo2024} and T2V-Zero~\cite{DBLP:conf/iccv/KhachatryanMTHW23} generate fully different video content to the anchor video frames, their performance for mTTA values is not satisfactory. With the consideration of temporal evolution, EQ-TAA (AVD) has superior performance in AUC and TTA$_{0.5}$ metrics to other EQ-TAAs. For the sunny daytime (A3D and CCD datasets), EQ-TAA (AVD) is more promising than other EQ-TAAs and achieves the best AUC value of $0.580$ for DADA-2000. 

\textbf{EQ-TAA vs. Supervised TAAs}: From the model architecture in Fig. \ref{fig4}, it can be seen that EQ-TAAs belong to the Fr. TAAs. At the same time, we differentiate them with different denotations in Table. \ref{tab2} and Table. \ref{tab3} for a clearer presentation. As for A3D and CCD datasets, EQ-TAAs seem not better than other supervised Ob.TAAs, which is because Ob.TAAs focus on targeted objects rather than frame-based EQ-TAAs. However, EQ-TAA (AVD) shows competitive performance with other object-centric ones in the DADA-2000 dataset even with a superior TTA$_{0.5}$ value to Ob. TAAs. DADA-2000 is more challenging than A3D and CCD because of severe weather and light conditions (as shown in Fig.\ref{fig5}). Therefore, object detection errors in DADA-2000 can be eliminated in the frame-centric TAAs, such as DRIVE \cite{DBLP:conf/iccv/Bao0K21} and  Cog-TAA \cite{CognitiveTAA}. Because of the frame background influence in the generated video frames, the video quality still needs to be enhanced compared with the real accident videos. Therefore, some supervised TAAs show better performance than EQ-TAAs. 

Notably, the AUC values of EQ-TAAs indicate that EQ-TAA (AVD) still needs space to be improved while EQ-TAA formulation is promising and does not need the laborious annotation of accident windows. From Fig.~\ref{fig7} and Fig.~\ref{fig8}, our accident video diffusion can be improved with the object-centric or causality inference.

\begin{figure}[!t]
\centering
\includegraphics[width=\hsize]{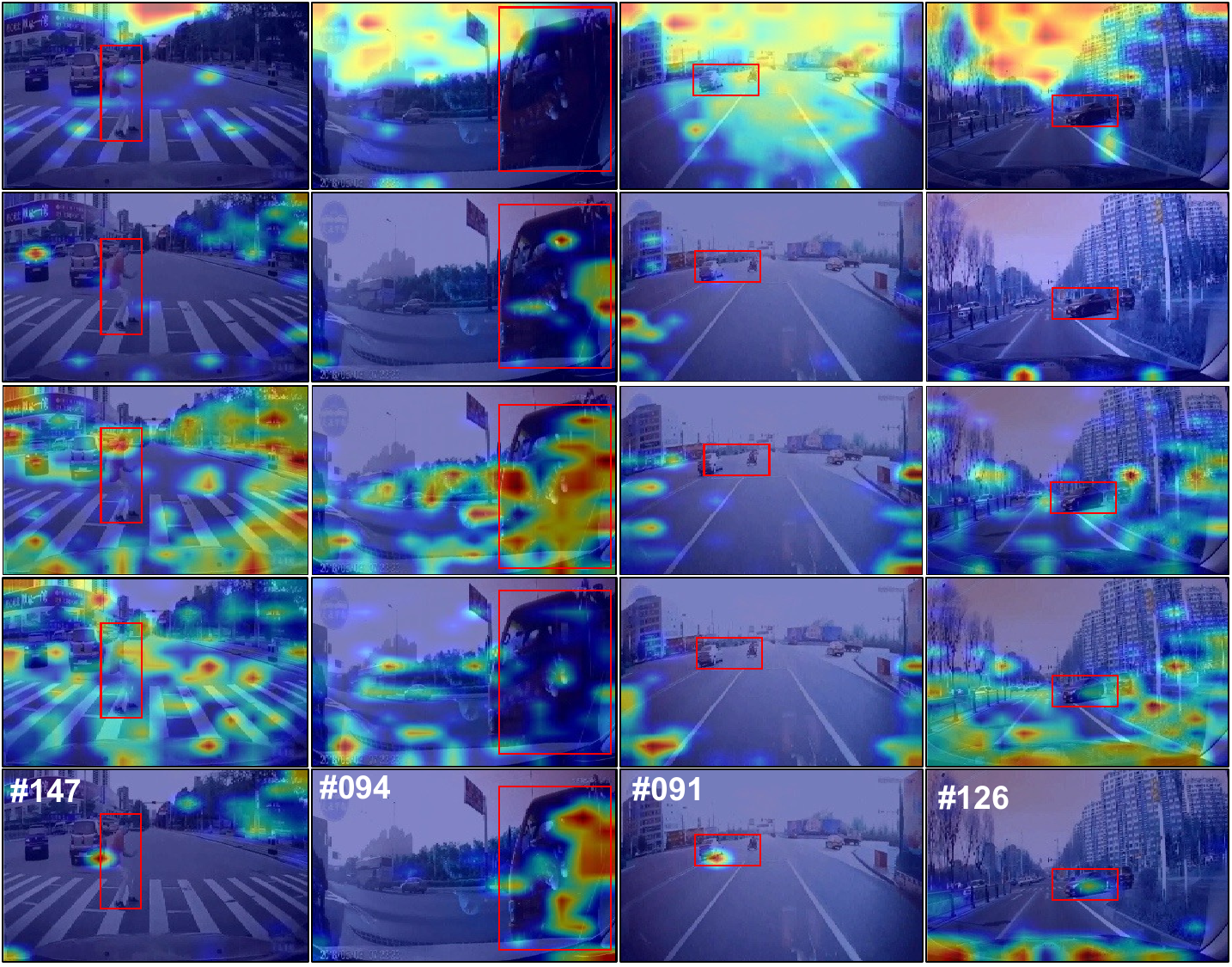}
\caption{The activated maps of the Layer Norm (LN) of ``S-Trans" by Grad-CAM++ \cite{chattopadhay2018grad}, where the EQ-TAA is trained by the training set obtained by Ho \emph{et al.} \cite{DBLP:journals/corr/abs-2204-03458} (the $1^{st}$ row), Tune-A-Video  \cite{DBLP:journals/corr/abs-2212-11565} (the $2^{nd}$ row), Text2Video-Zero~\cite{DBLP:conf/iccv/KhachatryanMTHW23} (the $3^{nd}$ row), ControlVideo~\cite{ControlVideo2024} (the $4^{nd}$ row), and our AVD (the $3^{rd}$ row),  respectively. The crashing objects are marked by red boxes.}
\label{fig9}
\end{figure}
 \begin{figure}
  \centering
 \includegraphics[width=\linewidth]{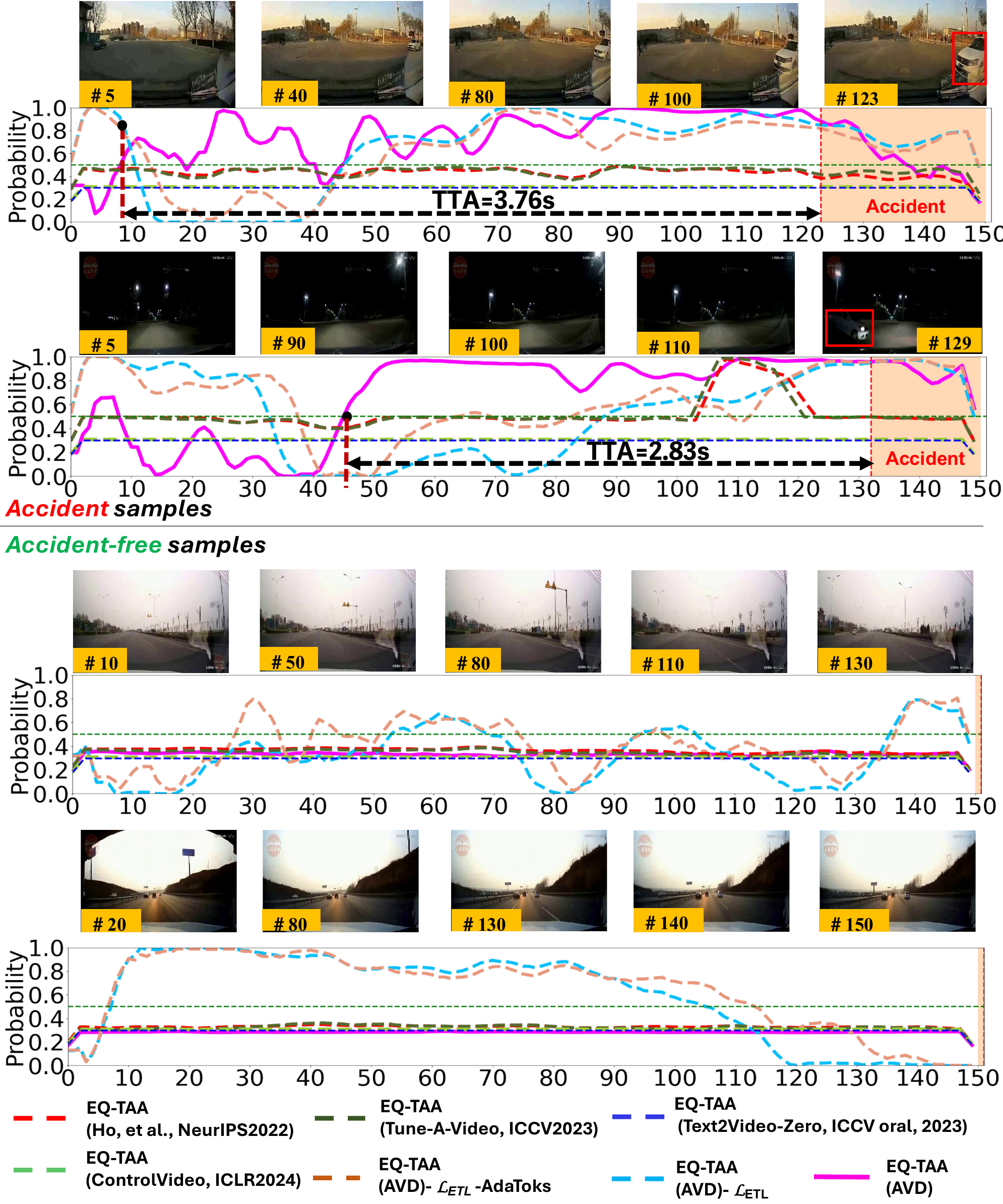}
\caption{The anticipated accident occurrence curves of two pairs of positive (top two rows) and negative (bottom two rows) accident videos, where one pair of real samples are sampled in the testing set of DADA-2000 \cite{DBLP:journals/tits/FangYQXY22}.}
  \label{fig10}
\end{figure}

We also visualize some snapshots in Fig. \ref{fig9}, where the activated map of the Layer Norm (LN) of S-Trans is generated by the Grad-CAM++ method \cite{chattopadhay2018grad}. From the results, it becomes clear that the activated maps by our AVD can localize the crashing object very well compared with other ones. The method of Ho \emph{et al.} \cite{DBLP:journals/corr/abs-2204-03458} always focuses on the sky regions. Tune-A-Video \cite{DBLP:journals/corr/abs-2212-11565} usually misses the crashing object. Because of the irrelevant video content in the generation, Text2Video-Zero~\cite{DBLP:conf/iccv/KhachatryanMTHW23} and ControlVideo~\cite{ControlVideo2024} focus on more background in TAA optimization, which provides the evidence to the low TTA values in Table.~\ref{tab2} and Table.~\ref{tab3}.

In summary, EQ-TAA (AVD) can outperform the other diffusion models on the A3D and CCD datasets, and obtain better AUC and TTA$_{0.5}$ values in the DADA-2000 dataset. For the supervised TAAs, the object-centric approach may be a good choice for clear environment conditions, while the frame-centric ones are better for severe conditions without detection error issues. 
Next, we evaluate the role of different components of EQ-TAA models.

\subsection{Ablations for EQ-TAA}
\subsubsection{EQ-TAA (AVD) vs. Other Supervised TAAs (AVD)}
To verify the performance of our EQ-TAA (described in Sec.~\ref{eq-taa}), we take the generated accident and accident-free video clips (\emph{i.e.}, 1800 pairs of clips) to train two supervised TAAs, \emph{i.e.}, XAI-AN \cite{DBLP:journals/corr/abs-2108-00273} and Cog-TAA \cite{CognitiveTAA}, and test the performance with aforementioned datasets. Notably, Cog-TAA \cite{CognitiveTAA} involves the scene description and driver attention to form a cognitive TAA. Because there is no driver attention and scene description in the generated video clips, we utilize the  Cog-TAA version only with video frame input. The training details follow the official setting of XAI-AN \cite{DBLP:journals/corr/abs-2108-00273} and Cog-TAA \cite{CognitiveTAA}. From the results in Table.~\ref{tab4}, Cog-TAA performs better in positive and negative classification than other methods for CCD and DADA-2000 datasets, but our EQ-TAA shows better performance on the earliness warning of accidents (larger mTTA values).
\begin{table}[!t]\small
\centering
\caption{The performance evaluation for different TAA methods \emph{driven by AVD}, where the best value of each metric is marked in \textbf{bold} font.}
\renewcommand{\arraystretch}{1.2}
     \setlength{\tabcolsep}{0.3mm}{
\begin{tabular}{l|cc|cc|cc}
\toprule[0.8pt]
 \multirow{2}[4]{*}{Method} & \multicolumn{2}{c|}{A3D \cite{DBLP:conf/iros/YaoXWCA19}}& \multicolumn{2}{c|}{CCD \cite{DBLP:conf/wacv/MallaCDCL23}}& \multicolumn{2}{c}{DADA-2000~\cite{DBLP:journals/tits/FangYQXY22}}\\
\cmidrule{2-7}      &AP$\uparrow$ &mTTA$\uparrow$ & AP$\uparrow$ &mTTA$\uparrow$& AP$\uparrow$ &mTTA$\uparrow$\\
\hline
XAI-AN \cite{DBLP:journals/corr/abs-2108-00273}&0.788&2.387&0.992&2.960&0.756&2.260\\
Cog-TAA \cite{CognitiveTAA} &0.905&4.128&\textbf{0.997}&4.041&\textbf{0.781}&4.063\\
EQ-TAA&\textbf{0.953}&\textbf{4.321} &0.989 &\textbf{4.407}&0.725&\textbf{4.258}\\
\toprule[0.8pt]
  \end{tabular}}
  \label{tab4}
  \end{table}
  
\subsubsection{Roles of ``ETL-Loss" and AdaToks}
In the EQ-TAA method, the special modules are the ``ETL-Loss" and the Adaptive Token Sampling (AdaToks). 
\begin{table}[!t]\footnotesize
  \centering
  \caption{The roles of $\mathcal{L}_{ETL}$ and ``AdaToks" in EQ-TAA (AVD).}
    \renewcommand{\arraystretch}{1.2}
        \setlength{\tabcolsep}{0.3mm}{
\begin{tabular}{lccccc|c}
\toprule[0.8pt]
Baselines & Datasets  & AP$\uparrow$ & AUC$\uparrow$ & TTA$_{0.5}$$\uparrow$ & mTTA$\uparrow$ & Var$\downarrow$ \\
\hline
EQ-TAA (AVD) &  & \textbf{0.725} & \textbf{0.580} & \textbf{4.338}  &  \textbf{4.498} &\textbf{0.049}\\
-$\mathcal{L}_{\text{ETL}}$ & DADA-2000 \cite{DBLP:journals/tits/FangYQXY22} & 0.702 & 0.523 &  3.894& 4.258&0.109 \\
-$\mathcal{L}_{\text{ETL}}$-AdaToks & & 0.724 & 0.438 & 3.933 & 4.378 &0.106  \\
\hline
EQ-TAA (AVD) & & \textbf{0.952} & - & \textbf{3.996}& \textbf{4.438}  &\textbf{0.084}\\
-$\mathcal{L}_{\text{ETL}}$ &A3D \cite{DBLP:conf/iros/YaoXWCA19}   & 0.905 & - & 3.995 & 4.437 &0.110 \\
-$\mathcal{L}_{\text{ETL}}$-AdaToks & & 0.905 & - &  3.596 &4.321 &0.107 \\
\hline
EQ-TAA (AVD) & & \textbf{0.989} & - & \textbf{4.402} & \textbf{4.875}  &\textbf{0.098} \\
-$\mathcal{L}_{\text{ETL}}$ & CCD \cite{DBLP:conf/wacv/MallaCDCL23}  & 0.982 & - & 4.401 & 4.845 &0.151  \\
-$\mathcal{L}_{\text{ETL}}$-AdaToks & & 0.982 & - & 4.401 &  4.407&0.154 \\
\toprule[0.8pt]
\end{tabular}}
\label{tab5}
\end{table}

To validate the effectiveness of ETL loss or AdaToks, we remove the ETL loss or AdaToks and compare it with the full EQ-TAA (AVD) model. From Table. \ref{tab5}, we observe a significant decrease in AP values for both the DADA-2000 and A3D datasets when $\mathcal{L}_{ETL}$ or AdaToks is absent. There is a significant degradation in TTA$_{0.5}$ or mTTA value when $\mathcal{L}_{ETL}$ or the AdaToks is removed. Larger TTA values mean earlier accident anticipation. The highest AUC value of our full EQ-TAA (AVD) in the DADA-2000 dataset seems to suggest that the anticipation with $\mathcal{L}_{ETL}$ or AdaToks may be more promising. In addition, we visualize some anticipated accident occurrence curves for real and synthetic videos in Fig. \ref{fig10}. We find that the anticipated accident score curves by ``EQ-TAA-$\mathcal{L}_{ETL}$" and ``EQ-TAA-$\mathcal{L}_{ETL}$-AdaToks" show frequent fluctuation. The TTA$_{0.5}$ metric utilized is precautious because it is computed by the time interval between \emph{the first warning time} (the first frame with the accident score larger than 0.5) with the starting time of the accident. Usually, the accident score may decrease to less than 0.5 after the first warning frame. However, to make a precautious warning, many works aforementioned still use it as a main metric. Therefore, we introduce a Variance (\textbf{Var}) metric computed by the frame-level variance of the accident scores ($>0.5$) to provide a prediction stability evaluation. The results of Table. \ref{tab5} show that the full EQ-TAA (AVD) generates a more stable prediction.
\subsubsection{Anticipation Stability of EQ-TAAs Driven by Different Diffusion Models}
\begin{figure}
  \centering
 \includegraphics[width=0.9\linewidth]{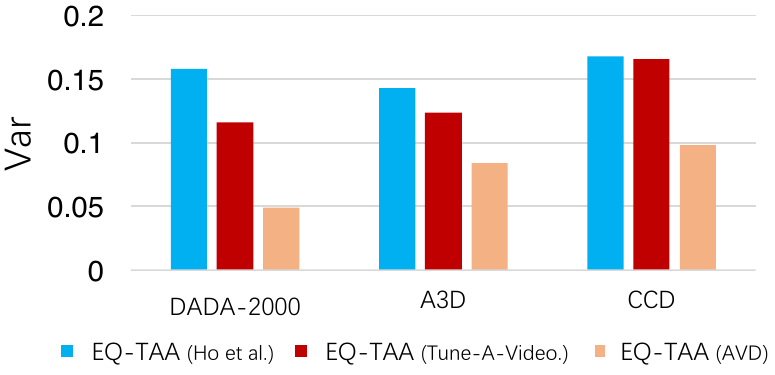}
\caption{The comparison of Var values for EQ-TAA framework driven by Ho \emph{et al.} \cite{DBLP:journals/corr/abs-2204-03458}, Tune-A-Video \cite{DBLP:journals/corr/abs-2212-11565}, and our AVD.}
  \label{fig11}
\end{figure}

In addition, we also check the prediction stability of full EQ-TAA driven by other video diffusion models. From the curves in Fig. \ref{fig10}, the full EQ-TAA model driven by AVD is more stable than ``EQ-TAA-$\mathcal{L}_{ETL}$" and ``EQ-TAA-$\mathcal{L}_{ETL}$-AdaToks". It indicates that the ETL loss and AdaToks modules are indispensable for stable anticipation. Besides, EQ-TAA (AVD) is better than EQ-TAA (Tune-A-Video), EQ-TAA (Ho \emph{et al.}), EQ-TAA (ControlVideo) and EQ-TAA (Text2Video-Zero). EQ-TAA (Ho \emph{et al.}), EQ-TAA (ControlVideo), and EQ-TAA (Text2Video-Zero) are powerless to determine the accident and accident-free videos because of the indiscriminative and irrelevant frame content aforementioned in Fig. \ref{fig7} and Fig. \ref{fig8}. Notably, EQ-TAA (ControlVideo), and EQ-TAA (Text2Video-Zero)  demonstrate failure for all the samples in Fig.~\ref{fig10}, which makes the \textbf{Var} value evaluation of them meaningless. As shown in Fig. \ref{fig11}, our EQ-TAA (AVD) is more stable for positive accident anticipation.

\subsubsection{Sample Scale vs. TAA Accuracy}
\begin{figure}[!t]
  \centering
 \includegraphics[width=\linewidth]{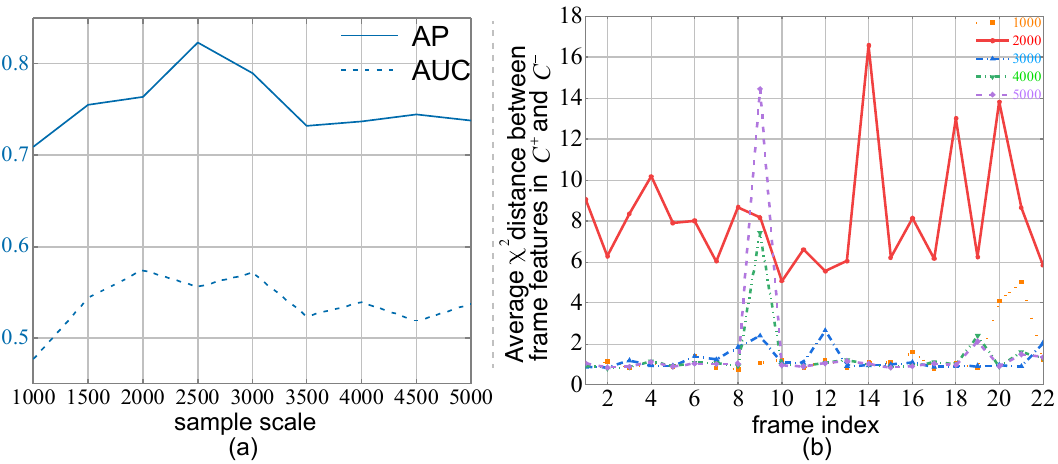}
\caption{Sample Scale vs. TAA Accuracy. (a) is accident anticipation accuracy (AP and AUC) concerning the sample scale, and (b) denotes the frame-wise feature difference for the generated positive and negative clips with 22 frames.}
  \label{fig12}
\end{figure}

As for video generation, it is necessary to investigate the sample scale of synthetic accident videos corresponding to the TAA performance. We check the AUC and AP values for different numbers of the generated samples by AVD in the TAA training phase. Fig. \ref{fig12} (a) shows that the AP and AUC values do not show a monotonic increase while demonstrating a decrease when the number of samples is more than 3000.  The reason is that our EQ-TAA takes the Video-to-Video (V2V) generation which leverages the anchor video clip $C^a$ to restrain the style of the generated ones (\emph{i.e.}, $C^+$ and $C^-$). When the sample scale increases, it may involve many similar video clips. We plot the average feature difference (Fig. \ref{fig12} (b)) of $C^+$ and $C^-$ with different sample scales by computing the $\chi^2$ distance of frame-wise feature vectors (22 frames) extracted by ResNet50. 2000 samples own the most manifest feature difference, which positively responds to the observation in Fig. \ref{fig12}(a), \emph{i.e.}, enlarging the feature difference of $C^+$ and $C^-$ is important.

\section{Conclusions and Discussions}
\label{con}

This work for the first time considers the text-video coherent alignment for accident scenarios in driving scenes, resulting in an Attentive Video Diffusion model (AVD) for handling the confounding issue of background frames for Traffic Accident Anticipation (TAA). The AVD generates smoother, style-consistent frames with anchor ones compared with other state-of-the-art video diffusion models. In addition, the AVD model facilitates the Equivariant TAA that can be trained by the video data collected from various driving scenes. EQ-TAA is verified to be comparable to state-of-the-art supervised traffic anticipation methods through extensive experiments. 

\textbf{Discussions}: Towards better accident video synthesis for the TAA task, compared with the overemphasized frame quality, the manifestation of the causal object is more important with the restriction of background influence and responsive change of text-guided video content. This is proved by the better positive and negative accident clip classification ability of models trained on real accident datasets. Therefore, the Inception Score (IS) metric is not good because it is easily influenced by the noisy context of the background even though the quality of generated images is good. As a video-based formulation of TAA tasks, the FVD metric is promising because of the temporal correlation. In addition, the metrics for evaluating the riskiness of generated video frames may be more appropriate and will be explored in the future.

Besides, more synthesized samples do not mean better TAA performance, which has two main factors: 1) the feature difference between the synthesized positive and negative samples, and 2) the diversity of the training samples. Better TAA performance prefers larger positive/negative feature differences and more diverse training samples. Therefore, if we sample the normal video clips as the anchor samples, it is better to enlarge the sampling interval of frames in one video or even from different videos.

{\small
\bibliographystyle{IEEEtran}
\bibliography{abbrev}
}

\end{document}